\documentclass[aps,pra,twocolumn,groupedaddress,floatfix,showpacs]{revtex4-2}
\usepackage{graphicx}
\usepackage{amsmath}
\usepackage{amssymb}
\usepackage{color}
\def\ket#1{|#1\rangle}
\def\bra#1{\langle#1|}

\begin{document}

\title{Collective dynamical Fermi suppression of optically induced inelastic scattering}

\author{Camen A. Royse, J. Huang, and J. E. Thomas}

\affiliation{Department of Physics, North Carolina State University, Raleigh, NC 27695, USA}

\date{\today}

\begin{abstract}
We observe strong dynamical suppression of optically induced loss in a weakly interacting Fermi gas as the $s$-wave scattering length is increased. The single, cigar-shaped cloud behaves as a large spin lattice in energy space with a tunable Heisenberg Hamiltonian. The loss suppression occurs as the lattice transitions into a magnetized state, where the fermionic nature of the atoms inhibits interactions. The data are quantitatively explained by incorporating spin-dependent loss into a quasi-classical collective spin vector model, the success of which enables the application of optical control of effective long-range interactions to this system.
\end{abstract}

\maketitle

%(mention s-wave scattering?)
In trapped, ultracold gases, understanding optically induced atom loss is essential for developing optical probes and control methods for
many-body systems~\cite{PhysRevLett.101.060406, Lapp_2019,skin_effect,Luo_PT,top_cont}.
Loss due to optically induced inelastic scattering has been used to study the BEC-BCS crossover in a Fermi gas via photoassociation~\cite{PhysRevLett.95.020404} and accompanies optical control of the $s$-wave scattering length~\cite{bauer,HaibinTwoField,OFR,Chin_opt,ArunEIT}.
Modeling optically induced two-body loss in a coherently prepared, weakly interacting Fermi gas is nontrivial, as it exhibits a coherent many-body spin evolution~\cite{DuSpinSeg2,LewensteinDynLongRange,SaeedPRASpinECorrel,Piechon,MuellerWeaklyInt,LaloeSpinReph,
ThywissenDynamicalPhases,KollerReySpinDep,WallEnergySpinLattice}.
Understanding this loss allows the spin dynamics to be probed and enables %the application of optical methods to
optical control of interactions in this system, %to this system,
which can be used to engineer the
Hamiltonian~\cite{SupportOnline}.

The Pauli principle plays an essential role in the evolution of the loss in an ultracold, weakly interacting Fermi gas, %as a pair of fermionic atoms
as the atoms cannot undergo  inelastic $s$-wave scattering when the spin state of a colliding atom pair is symmetric. This is especially relevant when
the gas evolves into a magnetized state, which occurs at a sufficiently large scattering length~\cite{ThywissenDynamicalPhases,JingjingCorrel}. Fermi gases have recently provided new demonstrations of the Pauli principle in degenerate samples, where Pauli blocking suppresses light scattering for %individual atoms
atoms in a Fermi sea~\cite{PauliKetterleHighDensity, PauliAtomLight,PauliLightScatt, PauliStimEm}.
In contrast, the suppression of light scattering reported here emerges from effective long range spin-spin interactions and is both dynamical and collective.

In this paper, we examine the collective suppression of optically induced inelastic scattering in a weakly interacting $^6$Li Fermi gas.
Each atom is prepared in a  pseudospin-state comprising a superposition of the two lowest hyperfine states %denoted by
$\ket{1}$ and $\ket{2}$. As the $s$-wave scattering length is increased, we observe a crossover from high to low optically induced loss. We develop and test a model for the spin-dependent loss, which shows that dynamical loss suppression arises from the onset of a magnetized state.

\begin{figure}[h]
  %  \hspace*{0.1 in}
  \includegraphics[width=3.40in]{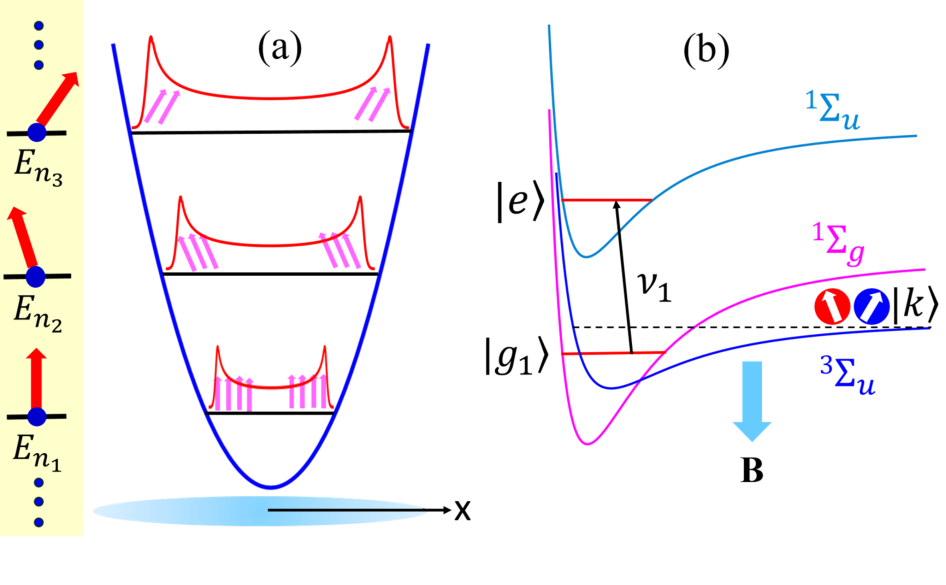}
    \caption{(a) Energy-space spin-lattice. Atoms remain fixed at  energy ``sites'' in a cigar-shaped optical trap (blue).
    Collective spin vectors (red arrows) are comprised of pseudospins in different transverse modes (pink arrows). %undergo an axial energy $E$-dependent evolution.
   Site-to-site couplings are determined by the overlap of the spatial probability distributions (orange). (b) Molecular states for two-body scattering near a Feshbach resonance. %A pair of atoms collides in a relative momentum state $\ket{k}$ in the magnetically-tunable electronic triplet channel $^3\Sigma_u$, which is hyperfine-coupled to a bound state $\ket{g_1}$ in the singlet channel $^1\Sigma_g$.
   Loss is induced by an optical field $\nu_1$ that drives a transition between $\ket{g_1}$ and  $\ket{e}$.}
    \label{fig:lattice}
\end{figure}

Tunable two-body scattering with optically induced loss is accomplished using a collisional (Feshbach) resonance,  Fig.~\ref{fig:lattice}(b). The resonance arises from hyperfine coupling between the triplet $^3\Sigma_u$ continuum $\ket{k}$ and a molecular vibrational state $\ket{g_1}$ in the singlet $^1\Sigma_g$ channel. At low temperatures, where $s$-wave scattering dominates, the $s$-wave scattering length $a_S$ is controlled by a bias magnetic field $B_z$, which tunes the total Zeeman-hyperfine energy of an incoming pair of atoms in state $\ket{k}$ near resonance with $\ket{g_1}$. Inelastic loss is induced by an optical field $\nu_1$ resonant with a transition from $\ket{g_1}$ to an excited electronic state $\ket{e}$, which spontaneously decays, causing loss of both atoms from the trap~\cite{HaibinTwoField,ArunEIT,NithyaSpatial}.
Related level schemes have been used for  optical control of $a_S$ via a $\nu_1$-induced light-shift of $\ket{g_1}$~\cite{bauer,HaibinTwoField,Chin_opt,ArunEIT}.
As the $s$-wave relative motion state is symmetric under the interchange of the atom labels, denoted $i,j$, scattering in the Fermi gas requires an antisymmetric two-atom hyperfine state, $\ket{\Psi_a(i,j)}=\frac{1}{\sqrt{2}}(\ket{1}_i\ket{2}_j-\ket{2}_i\ket{1}_j)$. Hence, the projection of the two-atom pseudo-spin state onto $\ket{\Psi_a(i,j)}$ determines the scattering probability.% in the energy lattice.

A ``weakly interacting'' Fermi gas is created by tuning $a_S$ to be small enough that the energy-changing collision rate $\propto a_S^2$ is negligible during each measurement period.
In the absence of optically induced loss, atoms remain fixed in their respective energy states, allowing the system to be described as a lattice in
a ``synthetic dimension''~\cite{synthLattice} formed by the energy eigenstates of the trapping potential, Fig.~\ref{fig:lattice}(a). Forward scattering between atoms at different energy ``sites''  causes rotations of the pseudospins, resulting in effective long-range couplings.
The lattice picture simplifies the description in comparison to a real space treatment, as the motional states of the atoms are fixed and the system evolves via pure spin dynamics, simulating a collective Heisenberg Hamiltonian~\cite{DuSpinSeg2,LewensteinDynLongRange,SaeedPRASpinECorrel,Piechon,MuellerWeaklyInt,LaloeSpinReph, ThywissenDynamicalPhases,KollerReySpinDep,WallEnergySpinLattice}.

In our experiments, the atoms are confined in a cigar-shaped optical trap.
The curvature of a bias magnetic field $\partial_x^2 B_z$ along the cigar axis $x$ produces a precession rate $\Omega_x'E$ for pseudospins of axial energy $E$.
Due to the tight transverse confinement,  $\Omega_y'$ and $\Omega_z'$ are 900 times smaller than $\Omega_x'$ and negligible.
This allows a 1D approximation for the lattice, where the spin-spin couplings between different transverse modes are replaced by a transverse mode-averaged coupling. Then, all pseudospins in a group with nearly the same axial energy $E$ evolve in the same way, as described by a \textit{collective} spin vector $\mathbf{S}(E,t)$ for each site.
We find that this model is in very good agreement with our observations~\cite{SaeedPRASpinECorrel,JingjingRewind,JingjingCorrel,SupportOnline}.

Without loss, the evolution of  $\mathbf{S}(E,t)$
is described by the spin Hamiltonian $H(E)=\boldsymbol{\omega}(E)\cdot\mathbf{S}(E)$, where
\begin{equation}
\boldsymbol{\omega}(E)=\Omega_x'E\,\hat{\mathbf{e}}_z+\sum_{E'\neq E}g(E,E')\,\mathbf{S}(E')\,.
\label{eq:omega}
\end{equation}
Here, $g(E,E')\propto a_S$ is the coupling between spins at axial energy sites $E$ and $E'\neq E$. In our experiments, the average coupling $\bar{g}\simeq 1.6$ Hz\,$\times\,2\pi$ for $a_S=5.0\,a_0$  and the rms spread in $\Omega_x'\,E$, denoted $\Omega_x'\sigma_E$, is $\simeq 1.4$ Hz$\,\times\,2\pi$.
Defining $\mathbf{S}(E,t)=S(E,t)\,\hat{\mathbf{S}}(E,t)$, where $\hat{\mathbf{S}}(E,t)$ is a unit vector,
\begin{equation}
\dot{\mathbf{S}}(E)=S(E)\,\dot{\hat{\mathbf{S}}}(E)+\dot{S}(E)\,\hat{\mathbf{S}}(E).
\label{eq:RotLoss}
\end{equation}
Here $S(E,t)=N_E(t)/2$  with $N_E(t)$ the number of atoms with axial energy $E$.
Neglecting loss, where $\dot{S}(E)=0$, the rotation of $\mathbf{S}(E)$, given by first term in Eq.~\ref{eq:RotLoss}, is determined by the Heisenberg equations,
\begin{equation}
\dot{\hat{\mathbf{S}}}(E,t)=\boldsymbol{\omega}(E,t)\times\hat{\mathbf{S}}(E,t).
\label{eq:rotation}
\end{equation}
We solve Eq.~\ref{eq:rotation} for the unit vectors $\hat{\mathbf{S}}(E,t)$ in a quasi-classical approximation, treating $\mathbf{S}(E,t)$ and $\mathbf{S}(E',t)$ as classical vectors.

\begin{figure}[htb]
\includegraphics[width=3.57in]{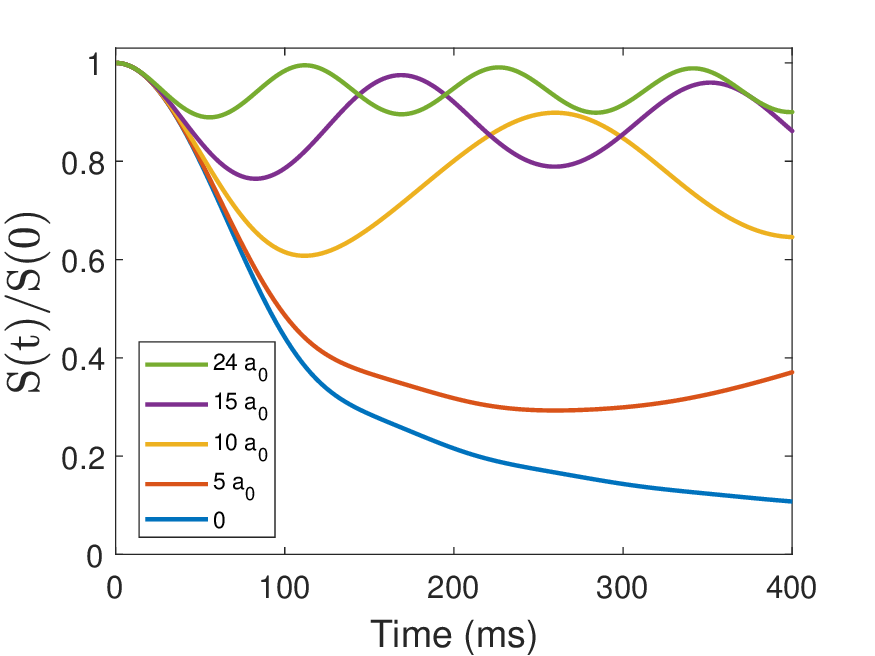}
\caption{Predicted magnitude of the total spin vector as a function of time for loss-free evolution with different $s$-wave scattering lengths. In the model, we use the experimental parameters given in the text. For $a_S/a_0=0,\,5,\,10,\,15,\,24$,  the respective interaction strengths are $\zeta=0,\,0.8, \,1.6, \,2.4, \,3.9$.}
\label{fig:TotalLength}
\end{figure}

The evolution of the collective spin vectors is determined by the competition between $\boldsymbol{\omega}_{\!B} (E)\equiv\Omega_x'E\,\hat{\mathbf{e}}_z$ and $\boldsymbol{\omega}_{\!M\!F}(E,t)\equiv\sum_{E'\neq E}g(E,E')\,\mathbf{S}(E',t)$ in Eq.~\ref{eq:omega}. As the lattice is not in thermal equilibrium, this competition results in two \textit{dynamical} phases: a spin-unlocked phase, where $\omega_B(E)$ dominates and a spin-locked phase, where $\omega_{MF}(E,t)\propto a_S$ dominates, independent of the sign of $a_S$. With increasing $|a_S|$, the lattice exhibits a crossover between these two dynamical phases~\cite{ThywissenDynamicalPhases,JingjingCorrel}. As the pseudo-spins are initially spin-polarized, they cannot interact until $\omega_B(E)$ causes the collective spin vectors to fan out with $E$-dependent angles in the transverse plane.  The crossover is characterized by  the interaction strength $\zeta\equiv \bar{g}/(\Omega'\sigma_E\sqrt{2})$. For small $a_S$, $\zeta$ is small and $\boldsymbol{\omega}_{\!B}(E)$ dominates, which is reflected in a low  magnitude of the total spin vector $S(t)=|\sum_E\mathbf{S}(E,t)|$, Fig. \ref{fig:TotalLength}.
%When $a_S$ is large enough that $\zeta > 1$,  spin-locking starts to occur.
We find that when $a_S$ is large enough that $\zeta \gtrsim 1.5$,
$\boldsymbol{\omega}_{\!M\!F}(E,t)$ dominates over  $\boldsymbol{\omega}_{\!B} (E)$ and the spins lock together. However, spin-locking suppresses scattering, enabling $\omega_B(E)$ to again fan out the spin vectors, which then re-enables scattering and subsequent spin locking, resulting in an oscillation of $S(t)$,  Fig. \ref{fig:TotalLength}. With increasing $\zeta\propto |a_S|$, the average $S(t)$ (magnetization) increases and the oscillation amplitude decreases.

Inelastic scattering is optically induced as described above, Fig.~\ref{fig:lattice}(b).
Spontaneous emission from $\ket{e}$ causes loss of both atoms, without heating or pumping into higher- or lower-energy trap modes, allowing use of the energy-space spin-lattice picture. With loss, the collective spin vectors rotate and change length, Eq.~\ref{eq:RotLoss} with $\dot{S}(E,t)=\dot{N}_E(t)/2\neq 0$. To incorporate loss into the model, we determine $N_E(t)$ as follows.

Loss due to two-body inelastic collisions between two species $A$ and $B$ with 3D densities $n_A(\mathbf{r},t)$ and $n_B(\mathbf{r},t)$ is generally %computed from the definition of the $AB$ inelastic cross section $\sigma_{inel}^{AB}$
modeled as
\begin{equation}
    \dot{n}_A(\mathbf{r},t)=\dot{n}_B(\mathbf{r},t)=-K_2^{AB} n_A(\mathbf{r},t)n_B(\mathbf{r},t)\,.
    \label{eq:two_body}
\end{equation}
Here $K_2^{AB}\equiv\langle v_r\sigma_{\rm inel}^{AB}\rangle$ with $\sigma_{\rm inel}^{AB}$ the $AB$ inelastic cross section and $\langle...\rangle$ denotes an average over relative speed $v_r$. In the energy-space spin-lattice, each energy corresponds to a definite spin vector. In our quasi-classical picture, atoms of axial energy $E$, in the spin state $\ket{\hat{\mathbf{S}}(E)}$, collide with atoms of energy $E'$ in the spin state $\ket{\hat{\mathbf{S}}(E')}$ for all $E'\neq E$.
To find $N_E(t)$, we generalize Eq.~\ref{eq:two_body} to model the loss of the spin-energy correlated 3D densities $n_E(\mathbf{r},t)$, i.e. the density of atoms with axial energy $E$:
\begin{eqnarray}
\dot{n}_E(\mathbf{r},t)=-\sum_{E'} K(E,E'\!,t)\,n_E(\mathbf{r},t)\,n_{E'}(\mathbf{r},t)\,,
\label{eq:densityEdecay}
\end{eqnarray}
where $K(E,E'\!,t)$ is the effective loss rate coefficient. Spin-dependent Fermi suppression is manifested in our expression for $K(E,E',t)$, which weights the two-body loss coefficient $K_2^a$ associated with the {\it anti-symmetric} two-atom hyperfine $\ket{\Psi_a(i,j)}$ by the probability that the incoming two-atom spin state $\ket{\hat{\mathbf{S}}(E)}_i\ket{\hat{\mathbf{S}}(E')}_j$ is in the state $\ket{\Psi_a(i,j)}$~\cite{SupportOnline},
\begin{eqnarray}
K(E,E'\!,t)=\frac{K_2^a}{4}[1-\hat{\mathbf{S}}(E,t)\cdot\hat{\mathbf{S}}(E',t)].
\end{eqnarray}
In the quasi-classical approximation, dynamical suppression of loss appears in the time dependence of the unit vectors, $\hat{\mathbf{S}}(E,t)$.
$K(E,E',t)$ has a maximum of $K_2^a/2$ when the colliding spin vectors are anti-parallel and vanishes when the spin vectors are parallel, which is most likely for a magnetized state.

To determine $N_E(t)$ from Eq.~\ref{eq:densityEdecay}, we employ a quasi-1D approximation, where  the 3D density factors~\cite{SupportOnline}: $n_E(\mathbf{r},t)= n_E(\rho,x,t)\simeq N_E(t)\mathcal{R}(\rho,t)|\phi_E(x)|^2$.
Here $x$ is the axial coordinate and $\rho$ is is the radial coordinate. We take $\mathcal{R}(\rho,t)$ to be the normalized transverse probability density, $\int d\rho\, 2\pi\rho\,\mathcal{R}(\rho,t)=1$ for all $t$ and $\int d^3\mathbf{r}\,n_E(\mathbf{r},t)=N_E(t)$. Integrals of Eq.~\ref{eq:densityEdecay} over $x$, $\rho$ result in coupled equations for $\dot{\mathcal{R}}(\rho,t)$ and $\dot{N}_E(t)$ ~\cite{SupportOnline}. Density-dependent loss causes $N_E(t)$ to decrease in time and $\mathcal{R}(\rho,t)$ to change shape, reducing the average transverse probability density $\bar{n}_\perp(t)=\int d\rho\, 2\pi\rho\,[\mathcal{R}(\rho,t)]^2$.
While we cannot directly measure $\mathcal{R}(\rho,t)$,
%We note that
including the time-dependence of  $\bar{n}_{\perp}(t)$ is essential, as is made apparent by comparing the measured loss rates to the model predictions with $\bar{n}_{\perp}=\bar{n}_{\perp}(t)$ and with $\bar{n}_{\perp}=\bar{n}_{\perp}(0)$~\cite{SupportOnline}. The evolution equations for $S(E,t)=N_E(t)/2$, $\hat{\mathbf{S}}(E,t)$, and $\mathcal{R}(\rho,t)$ determine the evolution of the total atom number $N(t)=\sum_E N_E(t)$.

To test the loss model, we measure the time-dependent loss of the total atom number $N(t)$  for scattering lengths $a_S=0$ to $24$ Bohr ($a_0$), corresponding to interaction strengths $\zeta= $0 to 5.39. The trapped gas is illuminated by a nominally uniform optical field resonant with the $\ket{g_1}\rightarrow \ket{e}$ transition and evolves for a variable amount of time before resonant absorption imaging of the atom densities for the spectrally resolved hyperfine states $\ket{1}$ and $\ket{2}$.

In the experiments, a gas of $N(0)=6\times 10^4$ $^6$Li atoms, is prepared in the weakly interacting regime~\cite{SaeedPRASpinECorrel}. The temperature of the gas is $T=0.18\,T_F$, where the Fermi temperature $T_F\simeq 0.75\,\mu$K. We use the calibration from Ref.~\cite{SaeedPRASpinECorrel} to tune to the desired scattering length $a_S(B)$, where RF spectroscopy precisely determines $B$.
A $0.5$ ms $\pi/2$ RF pulse is applied to a $z$-polarized sample to prepare the atoms in an equal superposition of the lowest-energy hyperfine states $\ket{1}$ and $\ket{2}$, i.e., the pseudospins are initially  polarized orthogonal to the magnetic field direction $z$.
Immediately following the pulse, a loss-inducing optical field is applied and the system evolves for a time $0\leq t\leq 400$ ms. The  Rabi frequency of the optical field is estimated to be~\cite{SupportOnline} $\Omega_1=\gamma_e/2$, where $\gamma_e=2\pi\times 11.8$ MHz is the spontaneous emission rate from the excited molecular state $\ket{e}$. Since the optical field is on resonance, there is no optical shift of the scattering length~\cite{NithyaSpatial}. The trap frequencies are $\omega_{\rho}=2\pi \times 668 $ Hz and $\omega_x=2\pi \times 25 $ Hz. A fit to  a zero-temperature Thomas-Fermi profile yields an axial width $\sigma_{TF}=330\,\mu$m. The radial width of 12 $\mu$m is computed from the ratio of the trap frequencies. For each measurement with a coherently prepared sample, the two-body loss rate coefficient $K_2^a$ is measured for a 50-50 mixture. These measured values of $K_2^a$ are used as inputs into the loss model.

\begin{figure}[htb]
\includegraphics[width=3.57in]{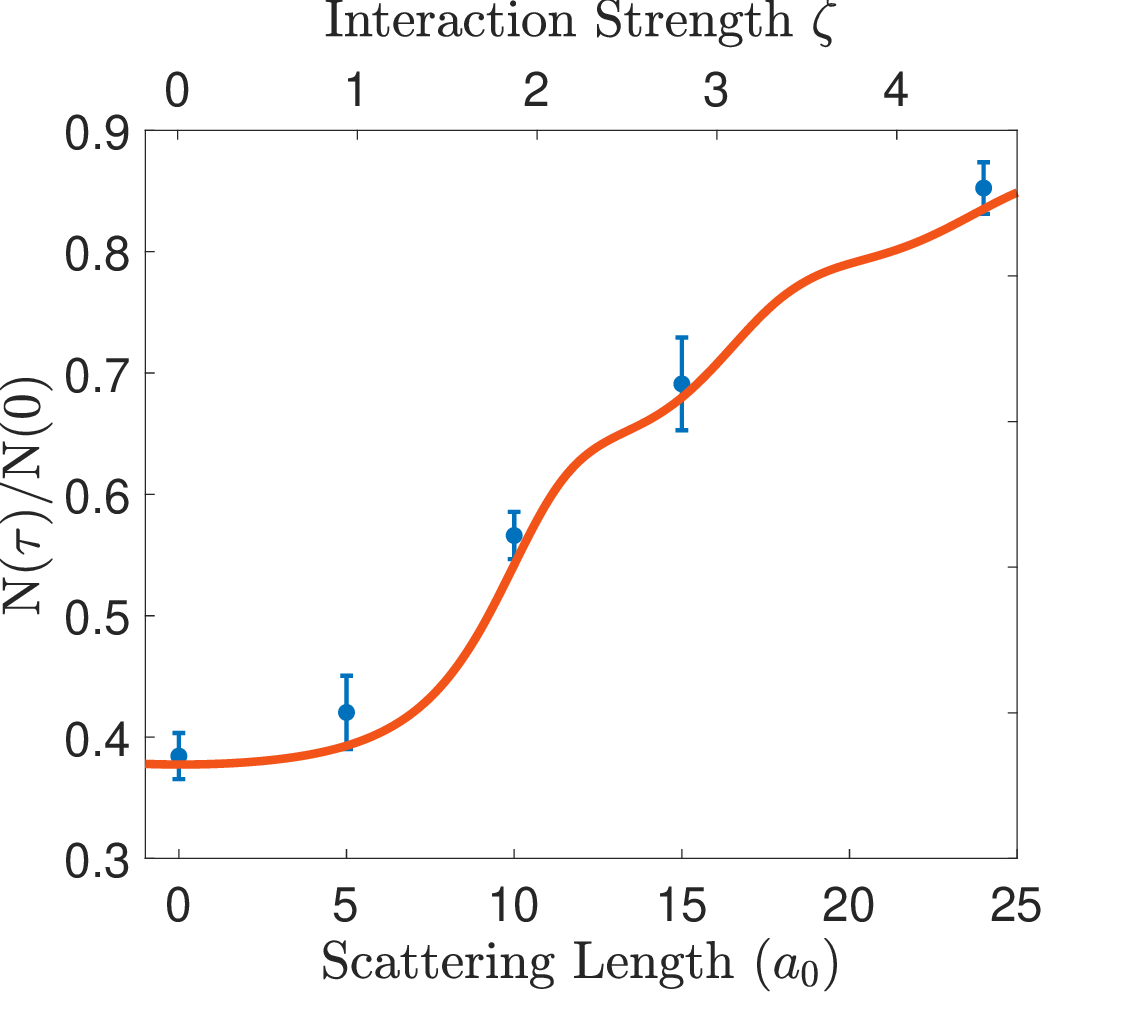}
\caption{Measurements of the atom fraction remaining after $\tau=$ 370 ms of illumination (blue points) vs scattering length and interaction strength $\zeta$, compared to the theoretical prediction (red curve). The densities and values of $K_2^a$ vary slightly for each measurement \cite{SupportOnline}.
For the prediction, we use the average values $N(0) = 6.1\times 10^4$ atoms,  $\sigma_{TF}=332\,\mu$m, and $K_2^a=62\, \mu$m$^3$/s.
\label{fig: transition}}
\end{figure}

\begin{figure*}[htb]
\begin{center}\
\hspace*{-0.25in}\includegraphics[width=5.5in]{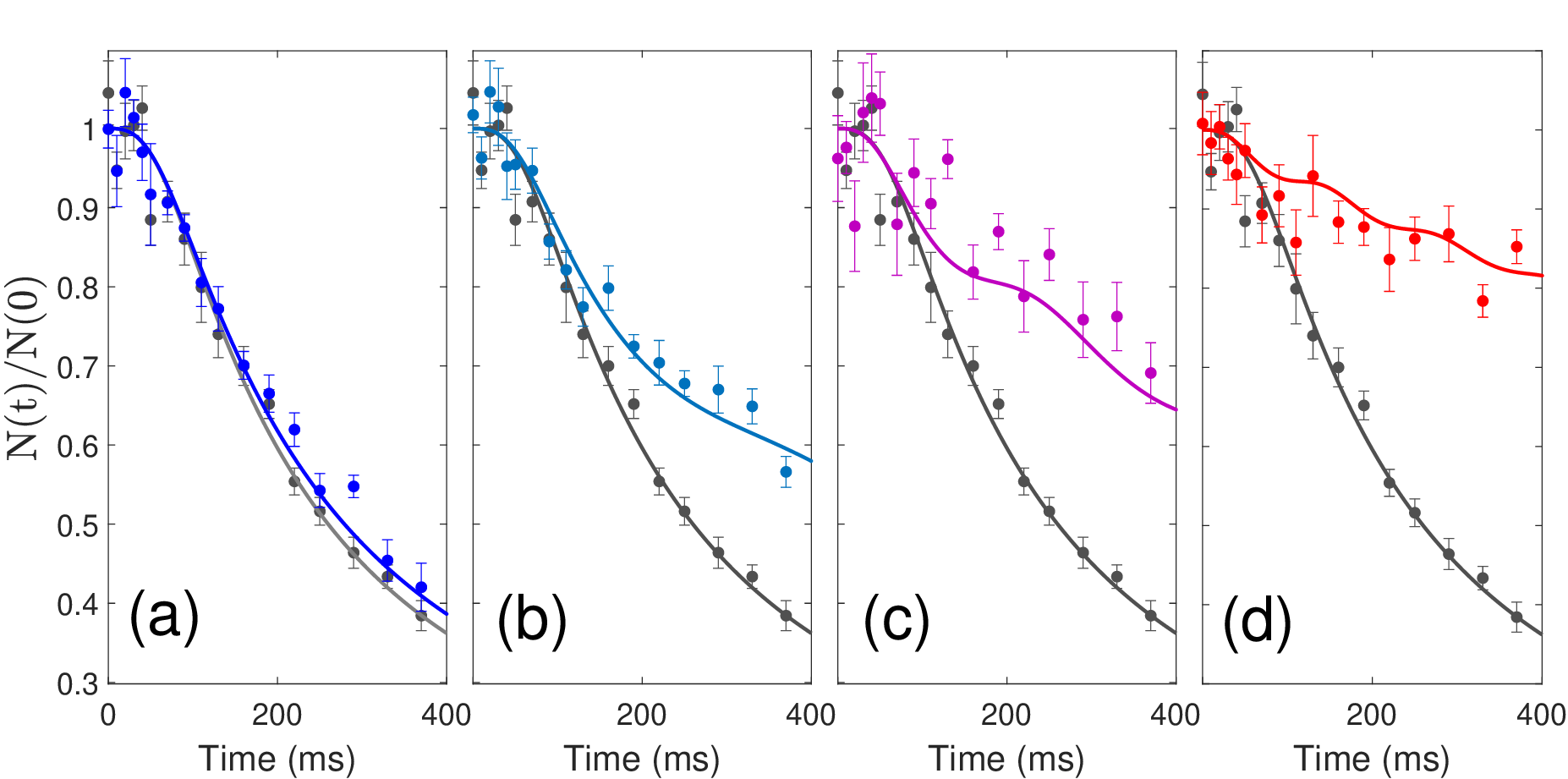}
\end{center}
\caption{Suppression of optically-induced loss versus illumination time as the scattering length is increased. $N(t)/N(0)$ is the atom fraction remaining after a time $t$. As a reference, the data and model for the non-interacting gas $a_{S}=0\,a_0$ ($\zeta=0$)  are shown in black on each plot. Each point represents the average of six shots, and the error bar is the standard deviation of the mean. (a) $a_{S}=5 \,a_0$ ($\zeta=1.03$), (b) $a_{S}=10\, a_0$ ($\zeta=2.32$), (c) $a_{S}=15\, a_0$ ($\zeta=3.59$), (d) $a_{S}=24 \,a_0$ ($\zeta=5.39$). Note that the interaction strength $\zeta$ is not precisely linear in the scattering length due to slight variations in the density. The measured values of $K_2^a$ and exact densities for each scattering length are given in \cite{SupportOnline}.}
\label{fig: LossvsScattLength}
\end{figure*}

The fraction of atoms remaining after 370 ms of illumination, $N(370\,\text{ms})/N(0)$, is shown in Fig. \ref{fig: transition} for the different scattering lengths. The data demonstrate a crossover between the unlocked and spin-locked dynamical phases, %the phase transition to a magnetized state,
where the Fermi suppression more than doubles the number of atoms remaining between the $a_S=\,$0 and 24 $a_0$ cases. Error bars represent the standard deviation of the mean for six shots. The prediction generated by the loss model (red curve) agrees well with the data. For the prediction, we use the averaged atom number, axial widths, and values of $K_2^a$ from the measurements.

Measurements of the fraction of atoms remaining throughout the evolution $N(t)/N(0)$ for coherently prepared samples are shown in Fig.~\ref{fig: LossvsScattLength}, along with the corresponding predictions using no free parameters. Predictions and measurements for $a_S=0\,a_0\,(\zeta=0)$, where interactions are absent, are shown as a reference, and agree very well.
The atom number is nearly stagnant for the first $\approx \,$80 ms, corresponding to the time needed for the energy-dependent Zeeman precession rates to separate the collective spin vectors. Once the spin vectors are sufficiently separated, the effective loss rate coefficient $K(E,E',t)$ becomes non-negligible and the atom number begins to decay.
At $a_S=5\,a_0$ $(\zeta=1.03)$, the data are almost indistinguishable from the $a_S=0\,a_0$ case, Fig.~\ref{fig: LossvsScattLength}a. This is consistent with Fig.~\ref{fig:TotalLength}, where, for $a_S=5\,a_0$  at our experimental densities,  the system is still in  the energy-dependent precession-dominated regime. The data show that a transition out of this dynamical phase occurs between $a_S=5\,a_0$ and $a_S=10\,a_0$ $(\zeta=2.32)$, where the measurements at $a_S=10\,a_0$ exhibit the onset of loss suppression, Fig.~\ref{fig: LossvsScattLength}b. The loss is further suppressed for the $a_S=15\,a_0 \,(\zeta=3.59)$ data, Fig.~\ref{fig: LossvsScattLength}c, and even more for the $a_S=24\,a_0\, (\zeta=5.39)$ data, Fig.~\ref{fig: LossvsScattLength}d, reflecting the increasing collective alignment of the spins, as depicted for the lossless case of Fig.~\ref{fig:TotalLength}.

Our collective spin vector model of loss for the energy-space lattice is in good quantitative agreement with measurements. The average of the values of $K_2^a$ used to generate the curves in Fig. \ref{fig: LossvsScattLength}, $\,62 \,\pm\, 6.2\,\mu\text{m}^3/s$, is in good agreement with the predicted value of $69.4\,\mu{\rm m}^3/s$ ~\cite{SupportOnline}.
For extraction of $K_2^a$ from loss measurements in a 50-50 mixture, we assume that a pair of colliding atoms is in the product state $\ket{1}_i\ket{2}_j$ and hence has a probability $|\langle \Psi_a(i,j) \ket{1}_i\ket{2}_j|^2=1/2$ to be in the antisymmetric spin state~\cite{SupportOnline}. However, we find that the values of $K_2^a$ used in the model need to be  {\it half} of those extracted from measurements in the 50-50 mixture. This origin of this discrepancy is not yet clear.

In summary, we have observed dynamical collective suppression of optically induced inelastic scattering in a coherently prepared, weakly interacting Fermi gas. As the scattering length is increased at fixed initial density, we observe a crossover from high to low loss. We understand this suppression via the Pauli principle, where the system undergoes a crossover into a magnetized dynamical phase with parallel collective spin vectors, Fig.~\ref{fig:TotalLength},  causing suppression of $s$-wave scattering. In this way, loss suppression serves as a new  probe of the magnetization of the system. We have developed a loss model that quantitatively agrees with observations and incorporates the many-body evolution of the collective spin vectors.
This work paves the way for tailoring of spin-spin couplings by optical control the interactions~\cite{SupportOnline}, as the accompanying loss can now be included in energy-space spin-lattice models.

Primary support for this research is provided by the Air Force Office of Scientific Research (FA9550-22-1-0329). Additional support is provided by the National Science Foundation (PHY-2006234 and PHY-2307107).

$^*$Corresponding author: jethoma7@ncsu.edu

%\bibliography{Fermigas4-17-arXiv}


\newpage
\widetext
\setcounter{figure}{0}
\setcounter{equation}{0}
\renewcommand{\thefigure}{S\arabic{figure}}
\renewcommand{\theequation}{S\arabic{equation}}

\appendix
\section{Supplemental Material}

In this supplemental material, we begin by reviewing the energy-space spin lattice model for the evolution of collective spin vectors in a weakly interacting Fermi gas without loss. Then we describe a generalized model, including two-body loss for the collective, energy-dependent spin vectors, which is compared to measurements in coherently prepared samples. Finally, we describe the experimental methods and the measurement of the optically-induced, two-body loss rate constant $K_2$ in two-state mixtures.

\subsection{Evolution of the Energy-Space Spin Lattice without Loss}
\label{sec:intro}

In the weakly interacting regime,  the $s$-wave scattering length $a_S$ is magnetically tuned to be sufficiently small that the collision rate is negligible over the experimental time scales. In this case, the energies of individual atoms are conserved, enabling an energy-space spin-lattice description~\cite{SaeedPRASpinECorrel,DuSpinSeg2}. For a cigar-shaped optical trap, the lattice can be approximated as one-dimensional, where the collective spin vectors $\mathbf{S}(E,t)$ are labeled by the axial energy $E$ for motion along the cigar axis, defined as $x$. The energy-space spin-lattice is described by a Heisenberg Hamiltonian, where the corresponding Heisenberg equations of motion yield an $E$-dependent rotation for the collective spin vector $\mathbf{S}(E,t)$ at each site.
In a mean-field description, the rotation arises from an effective site $E$-dependent Zeeman field and the effective magnetic field arising from the spin-spin coupling to all other sites $E'\neq E$. We employ a quasi-classical description, where $\mathbf{S}(E,t)$ is treated as a classical vector.

The experiments employ a $^6$Li Fermi gas in a superposition of the two lowest hyperfine-Zeeman states, which are denoted $\ket{1}\equiv\ket{\uparrow_z}$ and $\ket{2}\equiv\ket{\downarrow_z}$. The curvature of the applied bias magnetic field, $\partial^2_xB_z$, and the difference in the magnetic moments for the two hyperfine states produce a spin-dependent axial harmonic trap frequency and a corresponding $E$-dependent rotation rate about the applied magnetic field, which we denote by $\mathbf{\Omega_B}(E)=\Omega'E\,\hat{\mathbf{z}}$. The site-to-site coupling, denoted by $g(E,E')$, arises from forward s-wave scattering.

Taking $\mathbf{S}(E,t)= S(E,t)\,\hat{\mathbf{S}}(E,t)$, where $\hat{\mathbf{S}}(E,t)$ denotes a unit vector, we find
\begin{equation}
\label{eq:rotation}
\dot{\hat{\mathbf{S}}}(E,t)=\mathbf{\Omega_B}(E)\times\hat{\mathbf{S}}(E,t)+\sum_{E'}\,g(E,E')\,\mathbf{S}(E',t)\times \hat{\mathbf{S}}(E,t).
\end{equation}
Without loss, each $\mathbf{S}(E,t)$ evolves via rotation. In this case, the magnitudes $|\mathbf{S}(E,t)|\equiv S(E,t)=S(E,t=0)$ are conserved. There is some flexibility in the definition of $S(E)$, as Eq.~\ref{eq:rotation} is invariant under the scale transformation $S(E)\rightarrow c(E)\,S(E)$ and $g(E,E')\rightarrow g(E,E')/c(E)$. We choose $S(E,t=0)$ to be
\begin{equation}
\label{eq:initialmagnitude}
    S(E,t=0)= N_E/2.
\end{equation}
Here $N_E=N\,P(E)$ is the number of atoms in axial energy group $E$, with $N$ the total atom number and $P(E)$ the probability distribution. In the model, we take $P(E)$  to be a zero-temperature Thomas-Fermi distribution for near-degenerate samples; for higher temperatures, we employ a Boltzmann distribution.
The collective spin vectors begin their evolution after  a $\pi/2$ RF pulse coherently rotates the initially $z$-polarized sample, so that
\begin{equation}
\label{eq:initialdirection}
\hat{\mathbf{S}}(E,t=0)=\mathbf{\hat{x}'},
\end{equation}
where $\mathbf{\hat{x}'}$ is defined in the Bloch frame, orthogonal to $\mathbf{\hat{z}}$.

For our choice of $S(E,t=0)$ in Eq.~\ref{eq:initialmagnitude}, the site-to-site couplings $g(E,E')$ in Eq.~\ref{eq:rotation} are given by
\begin{equation}
\label{eq:couplings}
g(E,E')=-\bar{n}_{\perp}\frac{8\pi\hbar}{m}\, \int dx\,|\phi_E(x)|^2|\phi_{E'}(x)|^2 \,a_S
\end{equation}
where $\phi_E(x)$ is the axial trap eigenstate for energy $E$.  Note that optical control of interactions allows $a_S\rightarrow a_S(x,t)$, so that $g(E,E')$ may be tailored. In Eq. \ref{eq:couplings}, we have assumed that the single-particle probability density takes the form $\mathcal{R}(\rho)\,|\phi_E(x)|^2$, where $x$ is the axial coordinate, $\rho$ is the transverse radial coordinate, $\mathcal{R}(\rho)$ is the transverse probability density, and $\int d\rho \,2\pi\rho\,\mathcal{R}(\rho)=1$. The overlap integral is evaluated using a WKB approximation. For a harmonic trap,

\begin{equation}
\label{eq: integral}
\int dx\,|\phi_E(x)|^2|\phi_{E'}(x)|^2=\frac{2}{\pi^2}\sqrt{\frac{m\bar{\omega}_x^2}{2|E-E'|}}\,{\rm EllipticK}\left[-\frac{\text{min}(E,E')}{|E-E'|}\right],
\end{equation}
In Eq.~\ref{eq:couplings}, $\bar{n}_\perp$ is the average transverse probability density,
\begin{equation}
\label{eq:nperp}
\bar{n}_{\perp}\equiv\int d\rho\, 2\pi\rho \,\mathcal{R}^2(\rho).
\end{equation}

For lossless evolution, $\mathcal{R}(\rho)$ is time-independent. Assuming a zero-temperature Thomas-Fermi distribution,
\begin{equation}
\label{eq:RTF}
\mathcal{R}(\rho)=\frac{3}{\pi\sigma_{\rho}^2}\left(1-\frac{\rho^2}{\sigma_{\rho}^2}\right)^2
\end{equation}
we obtain $\bar{n}_{\perp}=9/(5\pi\sigma_{\rho}^2)$. For the Maxwell-Boltzmann distribution,
\begin{equation}
\label{eq:Rgaussian}
\mathcal{R}(\rho)=\frac{1}{\pi\sigma_{\rho}^2}\,e^{-\rho^2/\sigma_{\rho}^2},
\end{equation}
we find  $\bar{n}_{\perp}=1/(\pi\sigma_{\rho}^2)$.

\subsection{Modeling Two-Body Loss in the Energy-Space Spin lattice}
\label{sec:twobodyloss}

Inelastic interactions are induced in the energy-space spin lattice by illuminating the coherently prepared clouds with an optical field. In this section, we describe our model for the loss in this system due to these interactions.

We begin by describing the interaction process: For the magnetic fields of interest, a collision between a pair of $^6$Li atoms, one in each of the two lowest hyperfine spin states, occurs nominally in the triplet electronic potential (where ``triplet'' refers to the two-electron spin state). For $s$-wave scattering, where the relative motion state is symmetric in the interchange of the two atoms, the two-\textit{atom} hyperfine state is the antisymmetric state,
\begin{equation}
\ket{\Psi_a(1,2)}=\frac{1}{\sqrt{2}}\,\left(\ket{\uparrow_z}_1\,\ket{\downarrow_z}_2-\ket{\downarrow_z}_1\ket{\uparrow_z}_2\right)\simeq |1,-1; 1, 1\rangle.
\label{eq:antisymm}
\end{equation}
At high magnetic fields, as used in the experiments, $|1,-1; 1, 1\rangle$ is the dominant triplet state in the interior basis, i.e., the  total electronic spin state is $S=1,M_S=-1$, the total nuclear spin state is $I=1,M_I=1$. This triplet state has a large hyperfine coupling to the dominant singlet electronic state $S=0$~\cite{WuThomasEffRange}, denoted $\ket{g}$, which is in the $38^{\rm th}$ vibrational state of the singlet ground molecular potential, producing a broad Feshbach resonance at 832.2 G~\cite{JochimPreciseFeshbach}.  The difference between the magnetic moments of the singlet and triplet states enables magnetic tuning of the $s$-wave scattering length near the resonance. The applied optical field drives transitions from $\ket{g}$ to the $64^{\rm th}$ electronically-excited vibrational state in the electronic singlet molecular potential, denoted $\ket{e}$ \cite{HaibinTwoField,ArunEIT}. Spontaneous emission from $\ket{e}$ causes the interaction to be inelastic, and we assume that the emission results in loss of both atoms without transfer of atoms between energy states, so that the energy-space spin lattice model remains appropriate.

Loss due to two-body inelastic collisions between a particle of species $A$ and a particle of species $B$ is generally
modeled as
\begin{equation}
    \dot{n}_A(\mathbf{r},t)=\dot{n}_B(\mathbf{r},t)=-K_2^{AB} n_A(\mathbf{r},t)n_B(\mathbf{r},t)
    \label{eq:two_body}
\end{equation}
where $n_A(\mathbf{r},t)$ is the 3D density of species $A$ and $n_B(\mathbf{r},t)$ is the 3D density of species $B$. It is assumed that only $A$ and $B$ interact, and that each inelastic collisions causes both atoms to be lost. Eq. \ref{eq:two_body} follows from the definition of the inelastic cross section of
the $AB$ interaction $\sigma_{inel}^{AB}$ where $K_2^{AB}\equiv\langle v_{rel}\sigma_{inel}^{AB}\rangle$ (the brackets denote the average over the relative speeds $v_{rel}$). This will be our basis for constructing our loss model.

\subsubsection{Optically-Induced Loss in the Energy Lattice}

To treat loss in the energy-space spin lattice, we consider the atoms at each energy site $E$ to be a ``species'' in the context of Eq. \ref{eq:two_body}. We associate a 3D density $n_E(\mathbf{r},t)$ to the group of atoms with energy $E$ and a collective spin vector $\mathbf{S}(E,t)$, and sum the inelastic collision rates for atoms of energy with $E$ with atoms of energies $E'$ over all $E'\neq E$ to obtain
\begin{equation}
\label{eq:energyloss}
\dot{n}_E(\mathbf{r},t)=-\sum_{E'}K(E,E'\!,t)\,n_{E'}(\mathbf{r},t)\,n_E(\mathbf{r},t).
\end{equation}
Here the total density is $n(\mathbf{r},t)=\sum_E n_E(\mathbf{r},t)$ and $K(E,E'\!,t)$ is the effective energy-dependent two-body loss rate coefficient.

We obtain $K(E,E'\!,t)$ by computing the probability that the pair of atoms in energy groups $E$ and $E'$ are in the antisymmetric spin state $\ket{\Psi_a(1,2)}$. We assume that the spin of each atom of energy $E$ is polarized along $\mathbf{S}(E,t)$, corresponding to the spin state $\ket{\hat{\mathbf{S}}(E,t)}$. In this case, atoms of energies $E$ and $E'$ are in states with definite spin polarizations, so that we can assume the incoming spin state for a colliding pair of atoms with energies $E$ and $E'$ is  $\ket{\hat{\mathbf{S}}(E,t)}_1\,\ket{\hat{\mathbf{S}}(E',t)}_2$. The probability amplitude to be in the singlet state is then found by the inner product of this state with $\ket{\Psi_a(1,2)}$, so that
\begin{equation}
K(E,E'\!,t)=K_2^a\,|\bra{\Psi_a(1,2)}\hat{\mathbf{S}}(E,t)\rangle_1\,\ket{\hat{\mathbf{S}}(E',t)}_2|^2.
\label{eq:KEEprime}
\end{equation}
where $K_2^a$ is the loss constant associated with the antisymmetric two-atom spin state, given in Eq. \ref{eq:antisymm}. Suppressing the time dependence, the energy-dependent spin states take the form,
 \begin{eqnarray}
 \ket{\hat{\mathbf{S}}(E)}_1&=&e^{-i\phi_E/2}\cos(\theta_E/2)\,\ket{\!\!\uparrow_z}_1\,+\,e^{i\phi_E/2}\sin(\theta_E/2)\,\ket{\!\!\downarrow_z}_1\nonumber\\
 \ket{\hat{\mathbf{S}}(E')}_2&=&e^{-i\phi_E'/2}\cos(\theta_E'/2)\,\ket{\!\!\uparrow_z}_2\,+\,e^{i\phi_E'/2}\sin(\theta_E'/2)\,\ket{\!\!\downarrow_z}_2.
 \label{eq:spinstates}
\end{eqnarray}
A straightforward calculation gives
\begin{equation}
|\bra{\Psi_a(1,2)}\hat{\mathbf{S}}(E,t)\rangle_1\,\ket{\hat{\mathbf{S}}(E',t)}_2|^2=\frac{1}{4}\,\left[1-\cos\theta_E\cos\theta_{E'}-\sin\theta_E\sin\theta_{E'}\cos(\phi_E-\phi_{E'})\right],
\label{eq:effectiveK}
\end{equation}
or, in terms of the unit vectors and restoring the time dependence,
\begin{equation}
\label{eq:KE-Eprime}
K(E,E'\!,t)\equiv\frac{K_2^a}{4}\left[1-\hat{\mathbf{S}}(E,t)\cdot\hat{\mathbf{S}}(E',t)\right].
\end{equation}
As expected, when the collective spin vectors for energy groups $E$ and $E'$ vectors are parallel, the corresponding unit vectors $\hat{\mathbf{S}}(E,t)$ and $\hat{\mathbf{S}}(E',t)$ are parallel and there is no loss. In contrast, maximum loss occurs when the unit vectors are anti-parallel, $K(E,E',t)\rightarrow K_2^a/2$ .
The unit vectors $\hat{\mathbf{S}}(E,t)$ are found from  Eq.~\ref{eq:rotation}, with
$S(E,t)=N_E(t)/2$, where the atom number $N_E(t)$ is self-consistently determined from Eqs.~\ref{eq:energyloss}~and~\ref{eq:KE-Eprime}, as we now show.

We begin by assuming that the spin-energy correlated 3D densities $n_E(\mathbf{r},t)$ can be factored as
\begin{equation}
\label{eq:densityE}
n_E(\mathbf{r},t)=n_E(x,\rho,t) = N_E(t)\,\mathcal{R}(\rho,t)\,|\phi_E(x)|^2,
\end{equation}
where $x$ is the axial coordinate and $\rho$ the transverse coordinate. As observed in the experiments and shown in Fig.~\ref{fig:effectofdensitychange} below, for nonzero $K_2^a$, the increase in the loss rate with increasing 3D density reshapes the spatial profile. For this reason, we assume that both the atom number $N_E(t)$ in each energy group and the transverse probability density $\mathcal{R}(\rho,t)$ are functions of time. Further,
we assume that $\mathcal{R}(\rho,t)$ is independent of $E$, and take $\int d\rho \,2\pi\rho\,\mathcal{R}(\rho,t)=1$ for all $t$.  Using Eq.~\ref{eq:densityE}, the spatial integral of the total density, $n(\mathbf{r},t)=\sum_En_E(\mathbf{r},t)$  yields  total atom number,
\begin{equation}
N(t)=\sum_E N_E(t).
\label{eq:Ntotal}
\end{equation}

Using Eq.~\ref{eq:densityE} in Eq.~\ref{eq:energyloss} and integrating over $x$, we obtain
\begin{equation}
\frac{d}{dt}\left[N_E(t)\mathcal{R}(\rho,t)\right]=-\sum_{E'}\eta(E,E',t)\left[N_{E'}(t)\mathcal{R}(\rho,t)\right]\,\left[N_E(t)\mathcal{R}(\rho,t)\right],
\label{eq:NER}
\end{equation}
where
\begin{equation}
\label{eq:eta}
\eta(E,E'\!,t)\equiv K(E,E'\!,t)\int dx\,|\phi_E(x)|^2|\phi_{E'}(x)|^2.
\end{equation}
Integrating Eq.~\ref{eq:NER} over $\rho$ and using Eq.~\ref{eq:nperpt}, we find
\begin{equation}
\label{eq:IntRho}
\dot{N}_E(t)\int d\rho\, 2\pi\rho \,\mathcal{R}(\rho,t)+N_E(t)\frac{d}{dt}\int d\rho\, 2\pi\rho \,\mathcal{R}(\rho,t)=-\bar{n}_{\perp}(t)\,\sum_{E'}\eta(E,E'\!,t)\,N_{E'}(t)\,N_E(t),
\end{equation}
where $\bar{n}_{\perp}(t)$ is the time-dependent average transverse probability density
\begin{equation}
\label{eq:nperpt}
\bar{n}_{\perp}(t)\equiv\int d\rho\, 2\pi\rho \,\mathcal{R}^2(\rho,t).
\end{equation}
Since  $\int d\rho \,2\pi\rho\,\mathcal{R}(\rho,t)=1$, Eq.~\ref{eq:IntRho} immediately yields
\begin{equation}
\label{eq:numberEevol}
\dot{N}_E(t)=-\bar{n}_{\perp}(t)\,\sum_{E'}\eta(E,E'\!,t)\,N_{E'}(t)\,N_E(t).
\end{equation}

Next, we sum Eq.~\ref{eq:NER} over $E$ and use Eq.~\ref{eq:Ntotal} to obtain
\begin{equation}
\dot{N}(t)\,\mathcal{R}(\rho,t)+N(t)\,\dot{\mathcal{R}}(\rho,t)=-\mathcal{R}^2(\rho,t)\sum_E\sum_{E'}\eta(E,E'\!,t)\,N_{E'}(t)\,N_E(t)=\dot{N}(t)\,\frac{\mathcal{R}^2(\rho,t)}{\bar{n}_{\perp}(t)}.
\label{eq:Radial}
\end{equation}
Here, the right-hand side has been simplified by using the sum of Eq.~\ref{eq:numberEevol} over $E$ and Eq.~\ref{eq:Ntotal},
\begin{equation}
\dot{N}(t)=-\bar{n}_{\perp}(t)\,\sum_E\sum_{E'}\eta(E,E'\!,t)\,N_{E'}(t)\,N_E(t).
\label{eq:doublesumE}
\end{equation}
Hence, the radial probability distribution obeys
\begin{equation}
\label{eq:radialN}
\dot{\mathcal{R}}(\rho,t)=\frac{\dot{N}(t)}{N(t)}\left[\frac{\mathcal{R}^2(\rho,t)}{\bar{n}_{\perp}(t)}-\mathcal{R}(\rho,t)\right].
\end{equation}
Using Eq.~\ref{eq:nperpt}, one readily verifies that the integral of Eq.~\ref{eq:radialN} over $\rho$ vanishes, so that the total transverse probability remains normalized to $1$ for all $t$. Further,
the right hand side is $\propto\dot{N}(t)\,[\mathcal{R}(\rho,t)-\bar{n}_{\perp}(t)]$, where $\dot{N}(t)<0$ when $K_2^a\neq 0$. Hence, near the center of the cloud, where $\mathcal{R}(\rho,t)>\bar{n}_{\perp}(t)$, the probability density decreases in time, while in the wings, where $\mathcal{R}(\rho,t)<\bar{n}_{\perp}(t)$, the probability density increases in time. The net effect of the loss is to increase the effective width of $\mathcal{R}(\rho,t)$, while preserving the normalization.

\subsubsection{Optically-Induced Loss in a Mixture}

For the loss model described above, we require the loss constant $K_2^a$ associated with a pair of atoms in the antisymmetric two-atom spin state $\ket{\Psi_a(1,2)}$. To obtain $K_2^a$, we measure the loss in a 50-50 incoherent mixture of $\ket{\uparrow_z}$ and $\,\ket{\downarrow_z}$, for which the 50-50 ratio is maintained throughout the evolution, and extract the fraction of the loss constant associated with the state $\ket{\Psi_a(1,2)}$. Considering the mixture to be comprised of atoms in the $\ket{\uparrow_z}$ state and the $\ket{\downarrow_z}$ state, we define the 3D densities associated with each state $n_{\uparrow}(\mathbf{r},t)$ and $n_{\downarrow}(\mathbf{r},t)$ and apply Eq. \ref{eq:two_body} to obtain
\begin{equation}
    \dot{n}_{\uparrow}(\mathbf{r},t)=\dot{n}_{\downarrow}(\mathbf{r},t)=-K_2^{\uparrow\downarrow}n_{\uparrow}(\mathbf{r},t)n_{\downarrow}(\mathbf{r},t).
    \label{eq:updown}
\end{equation}
We assume that the incoming state is a product state $\ket{\uparrow_z}_1\ket{\downarrow_z}_2$. Then, the probability to be in the antisymmetric two-atom spin state is  $|\langle\Psi_a(1,2)\ket{\uparrow_z}_1\ket{\downarrow_z}_2|^2=1/2$,
\begin{equation}
    K_2^{\uparrow\downarrow}=K_2^{a}\times 1/2.
    \label{eq:Kupdown}
\end{equation}
With $n_{\uparrow}(\mathbf{r},t)+n_{\downarrow}(\mathbf{r},t)=n(\mathbf{r},t)$ the total density and $n_{\uparrow}(\mathbf{r},t)=n_{\downarrow}(\mathbf{r},t)=n(\mathbf{r},t)/2$ for a 50-50 mixture, Eq.~\ref{eq:updown}
yields
\begin{equation}
\dot{n}(\mathbf{r},t)=-\frac{1}{4}K_2^a\,n^2(\mathbf{r},t).
\label{eq:mixdecay}
\end{equation}
 Eq. \ref{eq:mixdecay}
may be solved analytically:
\begin{equation}
\label{eq:densitytime}
n(\mathbf{r},t)=\frac{n(\mathbf{r},0)}{1+\frac{1}{4}\,K^a_2n(\mathbf{r},0)\,t}.
\end{equation}
Integrating Eq.~\ref{eq:densitytime} over all three spatial dimensions, the total atom number $N(t)$ is predicted as a function of time, given $n(\mathbf{r},0)$:
\begin{equation}
\label{eq:numberevolmixture}
N(t)=\int\! dx\int\! 2\pi \rho\,d\rho\,\frac{n(\mathbf{r},0)}{1+\frac{1}{4}\,K_2^an(\mathbf{r},0)\,t}.
\end{equation}
To measure $K_2^a$, then, we fit measurements of the atom number $N(t)$ in the 50-50 mixture to Eq. \ref{eq:numberevolmixture}. This is further described in \S~\ref{sec:mixture}, where we show that $K_2^a$ is independent of the relative speed near the zero crossing of the broad Feshbach resonance in $^6$Li, see Eq.~\ref{eq:K2Calc}. However, as will also be discussed in \S~\ref{sec:mixture},  we must \textit{halve} the measured $K_2^a$ before inserting it into Eq. \ref{eq:effectiveK} in order to reach agreement with the loss measurements in the energy lattice.

\subsection{Evolution of the Energy-Space Spin Lattice with Loss}

To model the energy-space lattice with optically-induced loss, we employ Eqs.~\ref{eq:numberEevol}~and~\ref{eq:radialN}, together with Eq.~\ref{eq:rotation}. These equations determine the evolution of the density for each energy group, the transverse profile and therefore the total density and the total number in the presence of loss, which are compared with the measurements.

Including the $E$-dependent loss, the magnitudes of the collective spin vectors in Eq.~\ref{eq:rotation}, $S(E,t)= N_E(t)/2$,  decrease with time. The evolution of $\mathbf{S}(E,t)$ includes both a rotation of the unit vectors and a time-dependent magnitude,
\begin{equation}
\label{eq:vectorderiv}
    \dot{\mathbf{S}}(E,t)=S(E,t)\,\dot{\hat{\mathbf{S}}}(E,t)+\dot{S}(E,t)\,\hat{\mathbf{S}}(E,t).
\end{equation}
The unit vectors $\hat{\mathbf{S}}(E,t)$ evolve according to Eq.~\ref{eq:rotation}, while the decay of the magnitudes $S(E,t)$ is determined by Eq.~\ref{eq:numberEevol} with $N_E(t)=2\,S(E,t)$ and Eqs.~\ref{eq:eta}~and~\ref{eq:KE-Eprime},
\begin{equation}
\label{eq:SnumberE}
\dot{S}(E,t)=-\sum_{E'} \kappa(E,E'\!,t)\,\left[S(E,t)S(E',t)-\mathbf{S}(E,t)\cdot\mathbf{S}(E',t)\right].
\end{equation}
Here, the effective loss rate $\kappa(E,E'\!,t)$ is given by
\begin{equation}
\label{eq:kappa}
\kappa(E,E'\!,t)\equiv\frac{K_2^a}{2}\,\bar{n}_{\perp}(t)\int dx\,|\phi_E (x)|^2|\phi_{E'}(x)|^2.
\end{equation}
We discuss the measurement of $K_2^a$ for mixtures in \S~\ref{sec:mixture}.

We rewrite the evolution of the transverse probability density, Eq.~\ref{eq:radialN}, as
\begin{equation}
\label{eq:radialS}
\dot{\mathcal{R}}(\rho,t)=\frac{\dot{S}(t)}{S(t)}\left[\frac{\mathcal{R}^2(\rho,t)}{\bar{n}_{\perp}(t)}-\mathcal{R}(\rho,t)\right].
\end{equation}
Here we have defined $S(t)\equiv\sum_E S(E,t)=N(t)/2$.  Eq.~\ref{eq:nperpt} shows that the site-to-site couplings of Eq.~\ref{eq:couplings} become time dependent for $K_2^a\neq 0$, $g(E,E')\rightarrow g(E,E',t)$, while the decay of $S(E,t)$ reduces the rotation rate of the unit vectors by reducing the magnitude of the mean field.

Including loss, the evolution of the energy-dependent collective spin vectors is determined by Eq.~\ref{eq:vectorderiv}, using Eq.~\ref{eq:rotation} to describe the rotation of the unit vectors and Eqs.~\ref{eq:SnumberE},~\ref{eq:radialS},~and~\ref{eq:nperpt} to determine the decay of the magnitudes.
The collective spin vectors are initialized according to Eqs.~\ref{eq:initialmagnitude}~and~\ref{eq:initialdirection}.
The initial condition for the transverse probability density,  $\mathcal{R}(\rho,0)$, is given by Eq.~\ref{eq:RTF} for a Thomas-Fermi distribution and by Eq.~\ref{eq:Rgaussian} for a Maxwell-Boltzmann distribution.

\subsection{Experimental Methods}

\subsubsection{Measurement of Loss in the Energy-Space Spin Lattice}
\label{sec:expcoh}

To test the loss model, we measure the time-dependent decay of the total atom number $N(t)$ in a cigar-shaped optical trap comprising a single focused CO$_2$ laser beam. The measurements are obtained for scattering lengths $a_S=0\,a_0$ to $24\,a_0$  at nominally the same density. Starting from a $\hat{\mathbf{z}}$-polarized sample, we employ a $0.5$ ms $\pi/2$ RF pulse to prepare an initially $\mathbf{\hat{x}'}$-polarized sample as described in \S~\ref{sec:intro}.  Immediately following the RF pulse, the trapped gas is illuminated by a uniform optical field locked on-resonance with the singlet molecular $g\rightarrow e$ transition (see \S~\ref{sec:twobodyloss}) and evolves for a variable amount of time $t$ before absorption imaging of the atom densities for the  $\ket{\uparrow_z}$ and $\ket{\downarrow_z}$ states, which are spectrally resolved.

In the experiments, we begin by evaporatively cooling a 50-50 mixture of atoms in the two lowest hyperfine states $\ket{\uparrow}_z\equiv\ket{1}$ and $\ket{\downarrow}_z\equiv\ket{2}$ at the broad Feshbach resonance near 832.2 G~\cite{JochimPreciseFeshbach}. Following forced evaporation by lowering the trap depth, the trap depth is increased so that the radial trap frequency is $\omega_{\rho}=2\pi \times 668.0$ Hz.
To avoid the formation of Feshbach molecules while tuning to the weakly interacting region near 527 G,  the magnetic field is swept up to 1200 G and resonant light is applied to expel one spin state, leaving a $\hat{\mathbf{z}}$-polarized spin sample. The magnetic field is then swept to produce scattering length  $a_S(B)$ of interest near $527$ G. The calibration of Ref.~\cite{SaeedPRASpinECorrel} determines $a_S(B)$, where magnetic field is measured by RF spectroscopy.

\begin{figure}[htb]
    \centering
    \includegraphics[width=3.5in]{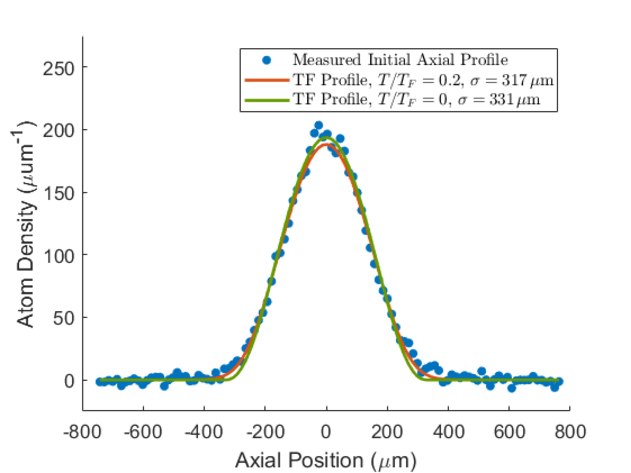}
    \caption{Thomas-Fermi fits to the sum of the initial axial profiles of the $\ket{\uparrow_z}$ and $\ket{\downarrow_z}$ states, immediately after the $\pi/2$ pulse, as used for measurement of $N(t)$ at 0 Bohr in the coherently-prepared sample. The fit of a finite-temperature 1D Thomas-Fermi profile yields the reduced temperature $T/T_F=0.2$. The fit is nearly identical to that of a zero-temperature 1D Thomas Fermi profile, justifying our use to a zero-temperature distribution in the model.}
    \label{fig:tempprofiles}
\end{figure}

After this preparation, the total number of atoms $N(0)\simeq 6.0\times 10^4$. A fit of the measured axial profile with a zero-temperature Thomas-Fermi distribution yields an axial width $\sigma_{TF}^x\simeq 331\,\mu$m, Fig.~\ref{fig:tempprofiles}. The radial width $\sigma_{TF}^{\rho}$ is computed from the ratio of trap transverse and axial frequencies, $\omega_{\rho}\sigma_{TF}^{\rho}=\omega_x\sigma_{TF}^x$. As noted in \S~\ref{sec:intro}, the curvature in the applied magnetic field results in a spin-dependent axial trapping force in the axial direction, where $\omega_{mag}=2\pi \times 16.3$ Hz. For the combined optical and magnetic trapping potentials near 527 G, the net axial trap frequency is measured to be $\omega_x=2\pi \times 25.0$ Hz. With $\omega_{\rho}=2\pi\times 668.0$ Hz, we find $\sigma_{TF}^{\rho}\simeq 12.0\,\mu$m.

To determine the temperature, we fit a 1-D finite-temperature Thomas-Fermi distribution to the initial axial profile.   The Fermi temperature $T_F$ for our harmonic trap is determined by
\begin{equation}
\label{eq:fermitemp}
E_F=k_BT_F=\hbar\,(6N\omega_\rho^2\omega_x)^{1/3}.
\end{equation}
Note that the 6 in Eq.~\ref{eq:fermitemp} reflects the fact that all $N$ atoms initially begin in an identical spin state. For the initial atom number and trap frequencies given in the last paragraph, we find $T_F\simeq 0.75\,\mu$K.  Using the {\it calculated} Thomas-Fermi radius $\sigma_{TF}=\sqrt{2E_F/(m\omega_x^2)}\simeq 317.0\,\mu$m, a fit to a finite-temperature Thomas-Fermi profiles yields  $T\simeq 0.20\,T_F$. Fig.~\ref{fig:tempprofiles} shows the averaged initial axial profile for a sample that is coherently prepared at 0 Bohr, along with the corresponding fitted finite-temperature 1D Thomas-Fermi and zero-temperature 1D Thomas-Fermi profiles.
The zero-temperature and finite-temperature Thomas-Fermi profiles are nearly identical, as expected for $T\simeq 0.20\,T_F$, justifying the use of an effective zero-temperature profile with a fitted width in the model.

For every scattering length, $K_2^a$ is measured from the loss in a 50-50 mixture, as discussed in \S~\ref{sec:mixture}.
Loss is induced by an optical beam propagating at an angle of $\simeq 49^\circ$ relative to the trap $x$-axis. The intensity half width at 1/e of the optical beam is $w=1.1$ mm, so that the projection of the full width of the optical beam at 1/e onto the cloud $x-axis$, is $2\,w\sin(49^\circ)\simeq 1.5\,w=1.6$ mm. This can be compared to the full width of the cloud $ 2\,\sigma_x\simeq 0.66$ mm. Hence, most of the atoms are illuminated near the peak intensity, $I=P/(\pi w^2)$.
The servo-stabilized beam  power is $7.6$ mW, so that  $I=2.0$ mW/mm$^2$. The Rabi frequency for the singlet electronic transition from the ground $38^{\rm th}$ vibrational state $\ket{g}$  to the excited $64^{\rm th}$ vibrational state has been measured~\cite{NithyaSpatial} to be $\Omega_1/2\pi = 4.4$ MHz\, $\sqrt{I{\rm [mW/mm}^2]}$. The Rabi frequency for the loss inducing beam is then $\Omega_1=0.53 \times \gamma_e$, where $\gamma_e=2\pi\times 11.8$ MHz is the rate of spontaneous emission from the excited molecular state~\cite{ArunEIT,HaibinTwoField}. The resonance frequency for each magnetic field value is found by finding the peak loss in the incoherent mixture as a function of frequency, which is prepared as described in \S~\ref{sec:mixture}.
Since the optical field is locked on resonance, there is no optical shift in the scattering length.

\begin{figure}[htb]
    \centering
    \includegraphics[width=3.5in]{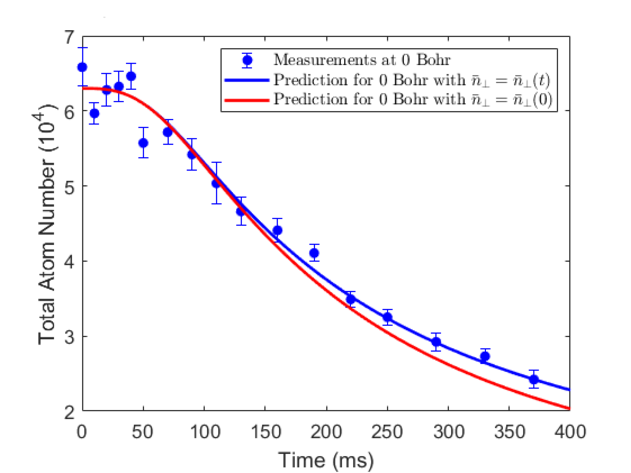}
    \caption{Predictions of loss in a coherently-prepared sample, with (blue) and without (red) the time dependence of the average transverse probability density $\bar{n}_{\perp}$. The time-dependence of $\bar{n}_{\perp}(t)$ arises from loss. Measurements for 0 Bohr in the coherently-prepared cloud are in  agreement with the model when $\bar{n}_{\perp}=\bar{n}_{\perp}(t)$ (blue). When $\bar{n}_{\perp}$ is taken to be constant (red), the tail of the loss curve does not agree with the measurements. For the 0 Bohr data, the  inputs into the loss model are $K_2^a=58.0\, \mu m^3/s$, the initial atom number $N=6.3\times 10^4$ and the width $\sigma_{TF}=331.0\,\mu$m.}
    \label{fig:effectofdensitychange}
\end{figure}

The importance of including the time dependence of $\mathcal{R}(\rho,t)$ in the model can be seen in the difference between the predictions for $\bar{n}_{\perp}=\bar{n}_{\perp}(0)$ and $\bar{n}_{\perp}=\bar{n}_{\perp}(t)$ at $a_S=0\,a_0$, as shown in Fig. \ref{fig:effectofdensitychange}. If $\bar{n}_{\perp}$ is taken to be constant, the model disagrees with the data for longer times. Accounting for the decrease in $\bar{n}_{\perp}(t)$ reduces the energy-dependent loss rate $\kappa(E,E'\!,t)$ of Eq. \ref{eq:kappa}, causing the tail of the loss curve to rise to match the data.

\subsubsection{Measurement of the Two-Body Loss Constant $K_2^a$ in a Mixture}
\label{sec:mixture}

To measure the two-body loss constant $K_2^a$, we measure the decay of the total number of atoms in an incoherent mixture of the $\ket{\uparrow_z}$ and $\ket{\downarrow_z}$ states. We employ a 50-50 mixture for which Eq.~\ref{eq:mixdecay} is valid, with
Eq.~\ref{eq:numberevolmixture} allowing $K_2^a$ to be determined from measurements of $N(t)$.
We model $n(\mathbf{r},0)$ as the Maxwell-Boltzmann distribution, which is appropriate for the higher temperature samples used in the mixture measurements,
\begin{equation}
\label{eq:MB}
n(\mathbf{r},0)=\frac{N(0)}{\pi\sigma_{\rho}^2\sigma_x\sqrt{\pi}}e^{-(\rho/\sigma_{\rho})^2-(x/\sigma_x)^2},
\end{equation}
with the axial size $\sigma_x$ determined from the measured spatial profiles,  the radial size $\sigma_{\rho}$ is found from the ratio of the trap frequencies. Using the initial density $n(\mathbf{r},0)$ in Eq.~\ref{eq:numberevolmixture},  the measured decay of the total number $N(t)$ determines $K^a_2$, which is used as a fit parameter. Here we expect that $K^a_2$ is independent of temperature, as discussed below (see Eq.~\ref{eq:K2Calc}).

\begin{figure}
    \centering
    \includegraphics[width=3.5in]{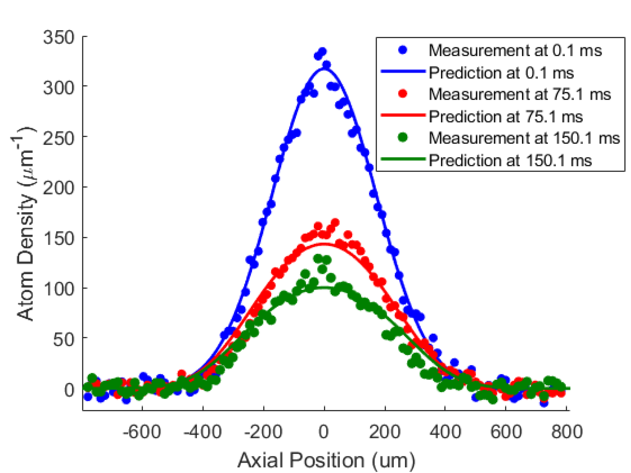}
    \caption{Measurements and predictions (Eq. \ref{eq: loss1D}) for the evolution of the axial profiles in a mixture. The magnetic field is tuned so that $a_S=15\,a_0$. The two-body loss rate constant  $K_2^a=2\times69.0\,\mu{\rm m}^3/s$ is determined from the fit of  $N(t)$, Eq.~\ref{eq:numberevolmixture} to the data.}
    \label{fig:mixdecayprofiles}
\end{figure}

To prepare the sample, a 50-50 incoherent mixture of atoms in spin states $\ket{\uparrow_z}$ and $\ket{\downarrow_z}$ undergoes forced evaporation at 300 G. Then the magnetic field is then swept upward to the magnetic field of interest in the weakly-interacting regime. This method avoids the formation of Feshbach molecules and subsequent loss. However, the efficiency of evaporation performed at 300 G, where the elastic scattering cross section is small, is reduced compared to that of the unitary gas at 832.2 G. For this reason, the samples used to measure $K_2^a$ are at a higher temperature than for the coherently prepared samples. We assume that $K_2^a$ is temperature independent, as $K_2^a$ is expected to exhibit a weak momentum dependence in the weakly interacting regime. At the magnetic field of interest, the loss-inducing optical field is applied, and the total number of atoms is measured as a function of time. The optical resonance frequency is determined by finding the peak loss point at each magnetic field of interest. Using the measured initial axial width and the initial radial width deduced from the ratio of the trap frequencies, the initial density profile is determined and Eq.~\ref{eq:mixdecay} is used to find $K_2^a$. This procedure is repeated for each scattering length $a_S$ employed in the experiments.

We determine the temperature from a fit of a Maxwell-Boltzmann distribution to the spatial profiles, $k_BT=m\omega_x^2\sigma_x^2/2$, where $\sigma_x$ is the fitted Gaussian width. For 15 $a_0$, this procedure gives $T = 0.56\,T_F$, where $T_F=0.79\,\mu$K is determined by
\begin{equation}
T_F=\frac{\hbar}{k_B}(3N\omega_{\rho}^2\omega_x )^{1/3}.
\end{equation}
Note that we have used a factor $3=6/2$ in place of the factor 6 in Eq.~\ref{eq:fermitemp}, as a 50-50 mixture has half of the total number of atoms $N$ in each spin state.

\begin{figure*}
    \centering
    \includegraphics[width=4.75in]{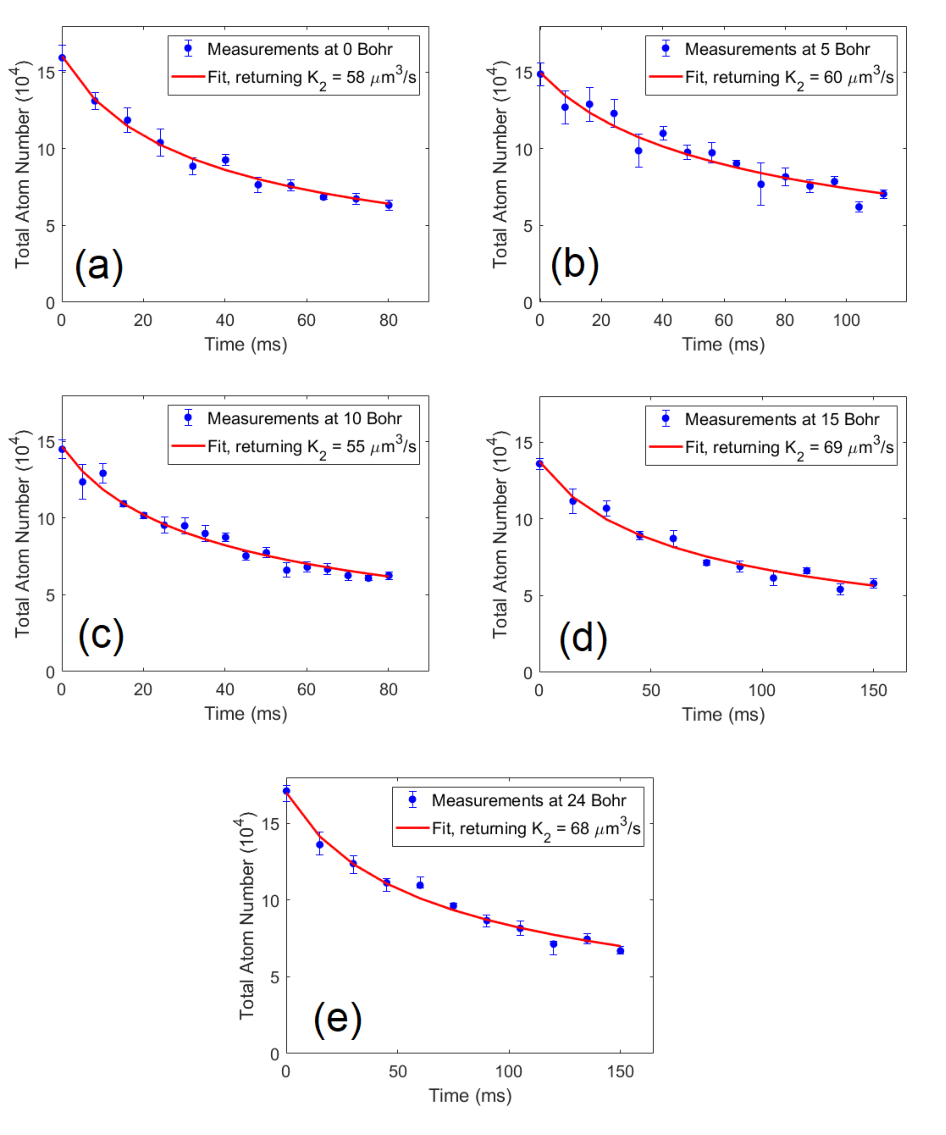}
   \caption{Measurements and predictions of loss in the 50-50 mixture at each scattering length. Different trap depths were used for 0 and 10 Bohr ($\omega_{\rho}=2\pi \times 1075 $ Hz, $\,\omega_{x}=2\pi \times 34 $ Hz) than for 5, 15, and 24 Bohr ($\omega_{\rho}=2\pi \times 675 $ Hz, $\,\omega_{x}=2\pi \times 23 $ Hz). (a) Loss at 0 Bohr, with an initial gaussian width 213 $\mu$m; (b) loss at 5 Bohr, with an initial gaussian width 241 $\mu$m; (c) loss at 10 Bohr, with an initial gaussian width 211 $\mu$m; (d) loss at 15 Bohr, with an initial gaussian width 243 $\mu$m; (e) loss at 24 Bohr, with an initial gaussian width 260 $\mu$m.
    \label{fig:loss}}
\end{figure*}

Measurements of $N(t)$ in 50-50 mixtures are shown in Fig.~\ref{fig:loss} for all of the scattering lengths of interest, using Eq.~\ref{eq:numberevolmixture} to determine $K_2^a$. The values extracted from the fit are displayed in Table~\ref{Table1}, where the uncertainty $\sigma_{K_2^a}$ is determined from the square root of the covariance matrix of the fit (note that this neglects the uncertainty in the initial density). The measured value of $K_2^a$ changes by $\simeq 10$\% as the scattering length is varied, most likely due to changes in the optical detuning and alignment from run-to-run. Note that the axial widths are smaller for the measurements at 0 and 10 $a_0$ than for 5, 15, and 24 $a_0$. The difference arises from the difference between the trap depths used for 0 and 10 $a_0$, where the trap frequencies were $\omega_{\rho}=2\pi \times 1075 $ Hz and $\,\omega_{x}=2\pi \times 34 $ Hz.  The 5, 15, and 24 $a_0$ data employed the smaller trap frequencies given in \S~\ref{sec:expcoh}. The faster timescales of loss for the 0 and 10 $a_0$ measurements reflect the higher density of the sample in the deeper trap.

Eq.~\ref{eq:densitytime} also predicts the time-dependent axial profiles $n_{1\text{D}}(x,t)$, which can be compared to measurements.  For the Maxwell-Boltzmann distribution of Eq.~\ref{eq:MB},
\begin{equation}
    n_{1\text{D}}(x,t)=\int d\rho\,2\pi\rho\,n(\mathbf{r,t})=\frac{4\pi\sigma_{\rho}^2}{K_2^a t}\,\text{ln}\left[1+\frac{K_2^a  t}{4\pi\sigma_{\rho}^2}\frac{N(0)}{\sigma_x\sqrt{\pi}}e^{-(x/\sigma_x)^2}\right].
    \label{eq: loss1D}
\end{equation}
In the limit $K_2^a t\rightarrow 0$, $n_{1\text{D}}(x,t)$ approaches a 1D gaussian distribution normalized to the initial total atom number $N(0)$, as it should. Using the $K_2^a$ determined from the fit to $N(t)$, we find that the predicted axial profiles are in quantitative agreement with the measured profiles, as shown for $a_S=15\,a_0$ in Fig.~\ref{fig:mixdecayprofiles}.

\begin{table}
\caption{Two-body loss coefficients.\label{Table1}}
\begin{center}
\begin{tabular}{|c | c | c |}
 \hline
 $a_S$ ($a_0$) & $K_2^a$ ($\mu$m$^3$/s) &  $\sigma_{K_2^a}$  ($\mu$m$^3$/s) \\
 \hline
 0 & 115 & 5 \\
 \hline
 5 &  120 & 11.6 \\
 \hline
 10 &  110 & 6.8 \\
 \hline
 15 &  138 & 10 \\
 \hline
 24 &  136 & 10 \\ [0.25ex]
 \hline
\end{tabular}
\end{center}
\end{table}

If the measurements in Table~\ref{Table1} are used in the energy-dependent loss rate coefficient $K(E,E',t)$ of Eq. \ref{eq:effectiveK}, however, the loss model does not agree with the measurements in the energy lattice. To obtain quantitative agreement between predictions and data for coherently prepared samples (as is shown in Fig. 2 and 3 in the main paper), we must divide the values of $K_2^a$ measured in the mixture by two. It is possible that we have incorrectly extracted $K_2^a$ by using Eq.~\ref{eq:Kupdown} or Eq.~\ref{eq:mixdecay}.

To gather evidence as to whether or not the factor of 1/2 is correct, we can compare the values of $K_2^a$ used to fit the coherently prepared sample in Fig.~3 of the main text to predictions for the optically-induced loss rate constant in $^6$Li. We take $K_2^{\rm calc}=-2\times\frac{8\pi\hbar}{m}\,a''$, where $a''<0$ is determined from the complex light-induced phase shift $\phi$ using $\tan\phi=- ik\,a''$ at the optical resonance. Here, $\hbar k$ is the relative momentum, and we assume $|k\,a''|<<1$ as is the case for our experiments. Note that a factor of two is included to be consistent with the antisymmetrized hyperfine state of Eq.~\ref{eq:antisymm} that defines $K_2^a$,  which in turn requires a symmetric spatial state with a total cross section~\cite{BraatenReview} $\sigma_{\rm tot}=8\pi/k\,Im\{f(0)\}$ and an elastic cross section $\sigma_{\rm el}=8\pi\,|f|^2$, with $f$ the s-wave scattering amplitude. The corresponding inelastic cross section $\sigma_{\rm inel}=\sigma_{\rm tot}-\sigma_{\rm el}=2\times\frac{\pi}{k^2}(1-|e^{2 i\phi}|^2)$ is {\it twice} that of Ref.~\cite{ArunEIT}, where the scattering atoms were treated as distinguishable and a factor $4\pi$ was used in the cross sections. The supplementary material of Ref.~\cite{ArunEIT} determines $a''$ using $x=k|a_{bg}|$ and $\tilde{\Delta}_0=(B-B_\infty)/\Delta B=-1$ in Eq.~S5, which gives $L(\tilde{\Delta}_0,x)\simeq 1$ in Eq.~S8 of Ref.~\cite{ArunEIT}, yielding
\begin{equation}
K^{\rm calc}_2=2\times\frac{8\pi\hbar\,|a_{bg}|}{m}\,\frac{\hbar\,\gamma_e}{4\,\mu_B\Delta B}\,\,\tilde{\Omega}_1^2\,,
\label{eq:K2Calc}
\end{equation}
where $\tilde{\Omega}_1\equiv\Omega_1/\gamma_e$.  With the parameters of Ref.~\cite{BartensteinFeshbach},  $a_{bg}=-1405\,a_0$, $\Delta B=300$ G, and $\mu_B/\hbar = 2\pi\times 1.4$ MHz/G, $\gamma_e=2\pi\times 11.8$ MHz we find $K^{\rm calc}_2=277.4\,\mu{\rm m}^3/s\,\,\tilde{\Omega}_1^2$, which gives $K^{\rm calc}_2=69.4\,\mu{\rm m}^3/s$ at $\tilde{\Omega}_1=0.5$ as used in the measurements. This result is in good agreement with the value $K_2^a= 69\,\mu{\rm m}^3/s$ that fits the decay of the coherently prepared sample at $15\,a_0$, but is, however, {\it half} the value $K_2^a= 138\,\mu{\rm m}^3/s$ extracted from measurements in the 50-50 mixture using Eq.~\ref{eq:mixdecay} as noted above. At present, we are unable to resolve this discrepancy, which may arise from applying Eq.~\ref{eq:mixdecay} to a very weakly interacting mixture or from an incorrect choice of the incoming two-atom state in deriving Eq.~\ref{eq:mixdecay}.

\begin{thebibliography}{33}%
\makeatletter
\providecommand \@ifxundefined [1]{%
 \@ifx{#1\undefined}
}%
\providecommand \@ifnum [1]{%
 \ifnum #1\expandafter \@firstoftwo
 \else \expandafter \@secondoftwo
 \fi
}%
\providecommand \@ifx [1]{%
 \ifx #1\expandafter \@firstoftwo
 \else \expandafter \@secondoftwo
 \fi
}%
\providecommand \natexlab [1]{#1}%
\providecommand \enquote  [1]{``#1''}%
\providecommand \bibnamefont  [1]{#1}%
\providecommand \bibfnamefont [1]{#1}%
\providecommand \citenamefont [1]{#1}%
\providecommand \href@noop [0]{\@secondoftwo}%
\providecommand \href [0]{\begingroup \@sanitize@url \@href}%
\providecommand \@href[1]{\@@startlink{#1}\@@href}%
\providecommand \@@href[1]{\endgroup#1\@@endlink}%
\providecommand \@sanitize@url [0]{\catcode `\\12\catcode `\$12\catcode
  `\&12\catcode `\#12\catcode `\^12\catcode `\_12\catcode `\%12\relax}%
\providecommand \@@startlink[1]{}%
\providecommand \@@endlink[0]{}%
\providecommand \url  [0]{\begingroup\@sanitize@url \@url }%
\providecommand \@url [1]{\endgroup\@href {#1}{\urlprefix }}%
\providecommand \urlprefix  [0]{URL }%
\providecommand \Eprint [0]{\href }%
\providecommand \doibase [0]{https://doi.org/}%
\providecommand \selectlanguage [0]{\@gobble}%
\providecommand \bibinfo  [0]{\@secondoftwo}%
\providecommand \bibfield  [0]{\@secondoftwo}%
\providecommand \translation [1]{[#1]}%
\providecommand \BibitemOpen [0]{}%
\providecommand \bibitemStop [0]{}%
\providecommand \bibitemNoStop [0]{.\EOS\space}%
\providecommand \EOS [0]{\spacefactor3000\relax}%
\providecommand \BibitemShut  [1]{\csname bibitem#1\endcsname}%
\let\auto@bib@innerbib\@empty
%</preamble>
\bibitem [{\citenamefont {Junker}\ \emph {et~al.}(2008)\citenamefont {Junker},
  \citenamefont {Dries}, \citenamefont {Welford}, \citenamefont {Hitchcock},
  \citenamefont {Chen},\ and\ \citenamefont {Hulet}}]{PhysRevLett.101.060406}%
  \BibitemOpen
  \bibfield  {author} {\bibinfo {author} {\bibfnamefont {M.}~\bibnamefont
  {Junker}}, \bibinfo {author} {\bibfnamefont {D.}~\bibnamefont {Dries}},
  \bibinfo {author} {\bibfnamefont {C.}~\bibnamefont {Welford}}, \bibinfo
  {author} {\bibfnamefont {J.}~\bibnamefont {Hitchcock}}, \bibinfo {author}
  {\bibfnamefont {Y.~P.}\ \bibnamefont {Chen}},\ and\ \bibinfo {author}
  {\bibfnamefont {R.~G.}\ \bibnamefont {Hulet}},\ }\bibfield  {title} {\bibinfo
  {title} {Photoassociation of a {Bose}-{Einstein} condensate near a {Feshbach}
  resonance},\ }\href {https://doi.org/10.1103/PhysRevLett.101.060406}
  {\bibfield  {journal} {\bibinfo  {journal} {Phys. Rev. Lett.}\ }\textbf
  {\bibinfo {volume} {101}},\ \bibinfo {pages} {060406} (\bibinfo {year}
  {2008})}\BibitemShut {NoStop}%
\bibitem [{\citenamefont {Lapp}\ \emph {et~al.}(2019)\citenamefont {Lapp},
  \citenamefont {Ang’ong’a}, \citenamefont {An},\ and\ \citenamefont
  {Gadway}}]{Lapp_2019}%
  \BibitemOpen
  \bibfield  {author} {\bibinfo {author} {\bibfnamefont {S.}~\bibnamefont
  {Lapp}}, \bibinfo {author} {\bibfnamefont {J.}~\bibnamefont {Ang’ong’a}},
  \bibinfo {author} {\bibfnamefont {F.~A.}\ \bibnamefont {An}},\ and\ \bibinfo
  {author} {\bibfnamefont {B.}~\bibnamefont {Gadway}},\ }\bibfield  {title}
  {\bibinfo {title} {Engineering tunable local loss in a synthetic lattice of
  momentum states},\ }\href {https://doi.org/10.1088/1367-2630/ab1147}
  {\bibfield  {journal} {\bibinfo  {journal} {New Journal of Physics}\ }\textbf
  {\bibinfo {volume} {21}},\ \bibinfo {pages} {045006} (\bibinfo {year}
  {2019})}\BibitemShut {NoStop}%
\bibitem [{\citenamefont {Zhou}\ \emph {et~al.}(2022)\citenamefont {Zhou},
  \citenamefont {Li}, \citenamefont {Yi},\ and\ \citenamefont
  {Cui}}]{skin_effect}%
  \BibitemOpen
  \bibfield  {author} {\bibinfo {author} {\bibfnamefont {L.}~\bibnamefont
  {Zhou}}, \bibinfo {author} {\bibfnamefont {H.}~\bibnamefont {Li}}, \bibinfo
  {author} {\bibfnamefont {W.}~\bibnamefont {Yi}},\ and\ \bibinfo {author}
  {\bibfnamefont {X.}~\bibnamefont {Cui}},\ }\bibfield  {title} {\bibinfo
  {title} {Engineering non-hermitian skin effect with band topology in
  ultracold gases},\ }\href@noop {} {\bibfield  {journal} {\bibinfo  {journal}
  {Communications Physics}\ }\textbf {\bibinfo {volume} {5}},\ \bibinfo {pages}
  {252} (\bibinfo {year} {2022})}\BibitemShut {NoStop}%
\bibitem [{\citenamefont {Li}\ \emph {et~al.}(2019)\citenamefont {Li},
  \citenamefont {Harter}, \citenamefont {Liu}, \citenamefont {de~Melo},
  \citenamefont {Joglekar},\ and\ \citenamefont {Luo}}]{Luo_PT}%
  \BibitemOpen
  \bibfield  {author} {\bibinfo {author} {\bibfnamefont {J.}~\bibnamefont
  {Li}}, \bibinfo {author} {\bibfnamefont {A.~K.}\ \bibnamefont {Harter}},
  \bibinfo {author} {\bibfnamefont {J.}~\bibnamefont {Liu}}, \bibinfo {author}
  {\bibfnamefont {L.}~\bibnamefont {de~Melo}}, \bibinfo {author} {\bibfnamefont
  {Y.~N.}\ \bibnamefont {Joglekar}},\ and\ \bibinfo {author} {\bibfnamefont
  {L.}~\bibnamefont {Luo}},\ }\bibfield  {title} {\bibinfo {title} {Observation
  of parity-time symmetry breaking transitions in a dissipative {Floquet}
  system of ultracold atoms},\ }\href@noop {} {\bibfield  {journal} {\bibinfo
  {journal} {Nature Communications}\ }\textbf {\bibinfo {volume} {10}},\
  \bibinfo {pages} {855} (\bibinfo {year} {2019})}\BibitemShut {NoStop}%
\bibitem [{\citenamefont {Ren}\ \emph {et~al.}(2022)\citenamefont {Ren},
  \citenamefont {Liu}, \citenamefont {Zhao}, \citenamefont {He}, \citenamefont
  {Pak}, \citenamefont {Li},\ and\ \citenamefont {Jo}}]{top_cont}%
  \BibitemOpen
  \bibfield  {author} {\bibinfo {author} {\bibfnamefont {Z.}~\bibnamefont
  {Ren}}, \bibinfo {author} {\bibfnamefont {D.}~\bibnamefont {Liu}}, \bibinfo
  {author} {\bibfnamefont {E.}~\bibnamefont {Zhao}}, \bibinfo {author}
  {\bibfnamefont {C.}~\bibnamefont {He}}, \bibinfo {author} {\bibfnamefont
  {K.~K.}\ \bibnamefont {Pak}}, \bibinfo {author} {\bibfnamefont
  {J.}~\bibnamefont {Li}},\ and\ \bibinfo {author} {\bibfnamefont {G.-B.}\
  \bibnamefont {Jo}},\ }\bibfield  {title} {\bibinfo {title} {Chiral control of
  quantum states in non-hermitian spin-orbit-coupled fermions},\ }\href@noop {}
  {\bibfield  {journal} {\bibinfo  {journal} {Nature Physics}\ }\textbf
  {\bibinfo {volume} {18}},\ \bibinfo {pages} {385} (\bibinfo {year}
  {2022})}\BibitemShut {NoStop}%
\bibitem [{\citenamefont {Partridge}\ \emph {et~al.}(2005)\citenamefont
  {Partridge}, \citenamefont {Strecker}, \citenamefont {Kamar}, \citenamefont
  {Jack},\ and\ \citenamefont {Hulet}}]{PhysRevLett.95.020404}%
  \BibitemOpen
  \bibfield  {author} {\bibinfo {author} {\bibfnamefont {G.~B.}\ \bibnamefont
  {Partridge}}, \bibinfo {author} {\bibfnamefont {K.~E.}\ \bibnamefont
  {Strecker}}, \bibinfo {author} {\bibfnamefont {R.~I.}\ \bibnamefont {Kamar}},
  \bibinfo {author} {\bibfnamefont {M.~W.}\ \bibnamefont {Jack}},\ and\
  \bibinfo {author} {\bibfnamefont {R.~G.}\ \bibnamefont {Hulet}},\ }\bibfield
  {title} {\bibinfo {title} {Molecular probe of pairing in the bec-bcs
  crossover},\ }\href {https://doi.org/10.1103/PhysRevLett.95.020404}
  {\bibfield  {journal} {\bibinfo  {journal} {Phys. Rev. Lett.}\ }\textbf
  {\bibinfo {volume} {95}},\ \bibinfo {pages} {020404} (\bibinfo {year}
  {2005})}\BibitemShut {NoStop}%
\bibitem [{\citenamefont {Bauer}\ \emph {et~al.}(2009)\citenamefont {Bauer},
  \citenamefont {Lettner}, \citenamefont {Vo}, \citenamefont {Rempe},\ and\
  \citenamefont {Dürr}}]{bauer}%
  \BibitemOpen
  \bibfield  {author} {\bibinfo {author} {\bibfnamefont {D.}~\bibnamefont
  {Bauer}}, \bibinfo {author} {\bibfnamefont {M.}~\bibnamefont {Lettner}},
  \bibinfo {author} {\bibfnamefont {C.}~\bibnamefont {Vo}}, \bibinfo {author}
  {\bibfnamefont {G.}~\bibnamefont {Rempe}},\ and\ \bibinfo {author}
  {\bibfnamefont {S.}~\bibnamefont {Dürr}},\ }\bibfield  {title} {\bibinfo
  {title} {Control of a magnetic {Feshbach} resonance with laser light},\
  }\href@noop {} {\bibfield  {journal} {\bibinfo  {journal} {Nature Physics}\
  }\textbf {\bibinfo {volume} {5}},\ \bibinfo {pages} {339–342} (\bibinfo
  {year} {2009})}\BibitemShut {NoStop}%
\bibitem [{\citenamefont {Wu}\ and\ \citenamefont
  {Thomas}(2012{\natexlab{a}})}]{HaibinTwoField}%
  \BibitemOpen
  \bibfield  {author} {\bibinfo {author} {\bibfnamefont {H.}~\bibnamefont
  {Wu}}\ and\ \bibinfo {author} {\bibfnamefont {J.~E.}\ \bibnamefont
  {Thomas}},\ }\bibfield  {title} {\bibinfo {title} {Optical control of
  {Feshbach} resonances in {Fermi} gases using molecular dark states},\
  }\href@noop {} {\bibfield  {journal} {\bibinfo  {journal} {Phys. Rev. Lett.}\
  }\textbf {\bibinfo {volume} {108}},\ \bibinfo {pages} {010401} (\bibinfo
  {year} {2012}{\natexlab{a}})}\BibitemShut {NoStop}%
\bibitem [{\citenamefont {Nicholson}\ \emph {et~al.}(2015)\citenamefont
  {Nicholson}, \citenamefont {Blatt}, \citenamefont {Bloom}, \citenamefont
  {Williams}, \citenamefont {Thomsen}, \citenamefont {Ye},\ and\ \citenamefont
  {Julienne}}]{OFR}%
  \BibitemOpen
  \bibfield  {author} {\bibinfo {author} {\bibfnamefont {T.~L.}\ \bibnamefont
  {Nicholson}}, \bibinfo {author} {\bibfnamefont {S.}~\bibnamefont {Blatt}},
  \bibinfo {author} {\bibfnamefont {B.~J.}\ \bibnamefont {Bloom}}, \bibinfo
  {author} {\bibfnamefont {J.~R.}\ \bibnamefont {Williams}}, \bibinfo {author}
  {\bibfnamefont {J.~W.}\ \bibnamefont {Thomsen}}, \bibinfo {author}
  {\bibfnamefont {J.}~\bibnamefont {Ye}},\ and\ \bibinfo {author}
  {\bibfnamefont {P.~S.}\ \bibnamefont {Julienne}},\ }\bibfield  {title}
  {\bibinfo {title} {Optical {Feshbach} resonances: Field-dressed theory and
  comparison with experiments},\ }\href
  {https://doi.org/10.1103/PhysRevA.92.022709} {\bibfield  {journal} {\bibinfo
  {journal} {Phys. Rev. A}\ }\textbf {\bibinfo {volume} {92}},\ \bibinfo
  {pages} {022709} (\bibinfo {year} {2015})}\BibitemShut {NoStop}%
\bibitem [{\citenamefont {Clark}\ \emph {et~al.}(2015)\citenamefont {Clark},
  \citenamefont {Ha}, \citenamefont {Xu},\ and\ \citenamefont
  {Chin}}]{Chin_opt}%
  \BibitemOpen
  \bibfield  {author} {\bibinfo {author} {\bibfnamefont {L.~W.}\ \bibnamefont
  {Clark}}, \bibinfo {author} {\bibfnamefont {L.-C.}\ \bibnamefont {Ha}},
  \bibinfo {author} {\bibfnamefont {C.-Y.}\ \bibnamefont {Xu}},\ and\ \bibinfo
  {author} {\bibfnamefont {C.}~\bibnamefont {Chin}},\ }\bibfield  {title}
  {\bibinfo {title} {Quantum dynamics with spatiotemporal control of
  interactions in a stable {Bose}-{Einstein} condensate},\ }\href
  {https://doi.org/10.1103/PhysRevLett.115.155301} {\bibfield  {journal}
  {\bibinfo  {journal} {Phys. Rev. Lett.}\ }\textbf {\bibinfo {volume} {115}},\
  \bibinfo {pages} {155301} (\bibinfo {year} {2015})}\BibitemShut {NoStop}%
\bibitem [{\citenamefont {Jagannathan}\ \emph {et~al.}(2016)\citenamefont
  {Jagannathan}, \citenamefont {Arunkumar}, \citenamefont {Joseph},\ and\
  \citenamefont {Thomas}}]{ArunEIT}%
  \BibitemOpen
  \bibfield  {author} {\bibinfo {author} {\bibfnamefont {A.}~\bibnamefont
  {Jagannathan}}, \bibinfo {author} {\bibfnamefont {N.}~\bibnamefont
  {Arunkumar}}, \bibinfo {author} {\bibfnamefont {J.~A.}\ \bibnamefont
  {Joseph}},\ and\ \bibinfo {author} {\bibfnamefont {J.~E.}\ \bibnamefont
  {Thomas}},\ }\bibfield  {title} {\bibinfo {title} {Optical control of
  magnetic {Feshbach} resonances by closed-channel electromagnetically induced
  transparency},\ }\href@noop {} {\bibfield  {journal} {\bibinfo  {journal}
  {Phys. Rev. Lett.}\ }\textbf {\bibinfo {volume} {116}},\ \bibinfo {pages}
  {075301} (\bibinfo {year} {2016})}\BibitemShut {NoStop}%
\bibitem [{\citenamefont {Du}\ \emph {et~al.}(2009)\citenamefont {Du},
  \citenamefont {Zhang}, \citenamefont {Petricka},\ and\ \citenamefont
  {Thomas}}]{DuSpinSeg2}%
  \BibitemOpen
  \bibfield  {author} {\bibinfo {author} {\bibfnamefont {X.}~\bibnamefont
  {Du}}, \bibinfo {author} {\bibfnamefont {Y.}~\bibnamefont {Zhang}}, \bibinfo
  {author} {\bibfnamefont {J.}~\bibnamefont {Petricka}},\ and\ \bibinfo
  {author} {\bibfnamefont {J.~E.}\ \bibnamefont {Thomas}},\ }\bibfield  {title}
  {\bibinfo {title} {Controlling spin current in a trapped {Fermi} gas},\
  }\href@noop {} {\bibfield  {journal} {\bibinfo  {journal} {Phys. Rev. Lett.}\
  }\textbf {\bibinfo {volume} {103}},\ \bibinfo {pages} {010401} (\bibinfo
  {year} {2009})}\BibitemShut {NoStop}%
\bibitem [{\citenamefont {Ebling}\ \emph {et~al.}(2011)\citenamefont {Ebling},
  \citenamefont {Eckardt},\ and\ \citenamefont
  {Lewenstein}}]{LewensteinDynLongRange}%
  \BibitemOpen
  \bibfield  {author} {\bibinfo {author} {\bibfnamefont {U.}~\bibnamefont
  {Ebling}}, \bibinfo {author} {\bibfnamefont {A.}~\bibnamefont {Eckardt}},\
  and\ \bibinfo {author} {\bibfnamefont {M.}~\bibnamefont {Lewenstein}},\
  }\bibfield  {title} {\bibinfo {title} {Spin segregation via dynamically
  induced long-range interactions in a system of ultracold fermions},\
  }\href@noop {} {\bibfield  {journal} {\bibinfo  {journal} {Phys. Rev. A}\
  }\textbf {\bibinfo {volume} {84}},\ \bibinfo {pages} {063607} (\bibinfo
  {year} {2011})}\BibitemShut {NoStop}%
\bibitem [{\citenamefont {Pegahan}\ \emph {et~al.}(2019)\citenamefont
  {Pegahan}, \citenamefont {Kangara}, \citenamefont {Arakelyan},\ and\
  \citenamefont {Thomas}}]{SaeedPRASpinECorrel}%
  \BibitemOpen
  \bibfield  {author} {\bibinfo {author} {\bibfnamefont {S.}~\bibnamefont
  {Pegahan}}, \bibinfo {author} {\bibfnamefont {J.}~\bibnamefont {Kangara}},
  \bibinfo {author} {\bibfnamefont {I.}~\bibnamefont {Arakelyan}},\ and\
  \bibinfo {author} {\bibfnamefont {J.~E.}\ \bibnamefont {Thomas}},\ }\bibfield
   {title} {\bibinfo {title} {Spin-energy correlation in degenerate weakly
  interacting {Fermi} gases},\ }\href@noop {} {\bibfield  {journal} {\bibinfo
  {journal} {Phys. Rev. A}\ }\textbf {\bibinfo {volume} {99}},\ \bibinfo
  {pages} {063620} (\bibinfo {year} {2019})}\BibitemShut {NoStop}%
\bibitem [{\citenamefont {Pi\'echon}\ \emph {et~al.}(2009)\citenamefont
  {Pi\'echon}, \citenamefont {Fuchs},\ and\ \citenamefont {Lalo\"e}}]{Piechon}%
  \BibitemOpen
  \bibfield  {author} {\bibinfo {author} {\bibfnamefont {F.}~\bibnamefont
  {Pi\'echon}}, \bibinfo {author} {\bibfnamefont {J.~N.}\ \bibnamefont
  {Fuchs}},\ and\ \bibinfo {author} {\bibfnamefont {F.}~\bibnamefont
  {Lalo\"e}},\ }\bibfield  {title} {\bibinfo {title} {Cumulative identical spin
  rotation effects in collisionless trapped atomic gases},\ }\href@noop {}
  {\bibfield  {journal} {\bibinfo  {journal} {Phys. Rev. Lett.}\ }\textbf
  {\bibinfo {volume} {102}},\ \bibinfo {pages} {215301} (\bibinfo {year}
  {2009})}\BibitemShut {NoStop}%
\bibitem [{\citenamefont {Natu}\ and\ \citenamefont
  {Mueller}(2009)}]{MuellerWeaklyInt}%
  \BibitemOpen
  \bibfield  {author} {\bibinfo {author} {\bibfnamefont {S.~S.}\ \bibnamefont
  {Natu}}\ and\ \bibinfo {author} {\bibfnamefont {E.~J.}\ \bibnamefont
  {Mueller}},\ }\bibfield  {title} {\bibinfo {title} {Anomalous spin
  segregation in a weakly interacting two-component {Fermi} gas},\ }\href
  {https://doi.org/10.1103/PhysRevA.79.051601} {\bibfield  {journal} {\bibinfo
  {journal} {Phys. Rev. A}\ }\textbf {\bibinfo {volume} {79}},\ \bibinfo
  {pages} {051601} (\bibinfo {year} {2009})}\BibitemShut {NoStop}%
\bibitem [{\citenamefont {Deutsch}\ \emph {et~al.}(2010)\citenamefont
  {Deutsch}, \citenamefont {Ramirez-Martinez}, \citenamefont {Lacro\^ute},
  \citenamefont {Reinhard}, \citenamefont {Schneider}, \citenamefont {Fuchs},
  \citenamefont {{Pi\'echon}}, \citenamefont {{Lalo\"{e}}}, \citenamefont
  {Reichel},\ and\ \citenamefont {Rosenbusch}}]{LaloeSpinReph}%
  \BibitemOpen
  \bibfield  {author} {\bibinfo {author} {\bibfnamefont {C.}~\bibnamefont
  {Deutsch}}, \bibinfo {author} {\bibfnamefont {F.}~\bibnamefont
  {Ramirez-Martinez}}, \bibinfo {author} {\bibfnamefont {C.}~\bibnamefont
  {Lacro\^ute}}, \bibinfo {author} {\bibfnamefont {F.}~\bibnamefont
  {Reinhard}}, \bibinfo {author} {\bibfnamefont {T.}~\bibnamefont {Schneider}},
  \bibinfo {author} {\bibfnamefont {J.~N.}\ \bibnamefont {Fuchs}}, \bibinfo
  {author} {\bibfnamefont {F.}~\bibnamefont {{Pi\'echon}}}, \bibinfo {author}
  {\bibfnamefont {F.}~\bibnamefont {{Lalo\"{e}}}}, \bibinfo {author}
  {\bibfnamefont {J.}~\bibnamefont {Reichel}},\ and\ \bibinfo {author}
  {\bibfnamefont {P.}~\bibnamefont {Rosenbusch}},\ }\bibfield  {title}
  {\bibinfo {title} {Spin self-rephasing and very long coherence times in a
  trapped atomic ensemble},\ }\href@noop {} {\bibfield  {journal} {\bibinfo
  {journal} {Phys. Rev. Lett.}\ }\textbf {\bibinfo {volume} {105}},\ \bibinfo
  {pages} {020401} (\bibinfo {year} {2010})}\BibitemShut {NoStop}%
\bibitem [{\citenamefont {Smale}\ \emph {et~al.}(2019)\citenamefont {Smale},
  \citenamefont {He}, \citenamefont {Olsen}, \citenamefont {Jackson},
  \citenamefont {Sharum}, \citenamefont {Trotzky}, \citenamefont {Marino},
  \citenamefont {Rey},\ and\ \citenamefont
  {Thywissen}}]{ThywissenDynamicalPhases}%
  \BibitemOpen
  \bibfield  {author} {\bibinfo {author} {\bibfnamefont {S.}~\bibnamefont
  {Smale}}, \bibinfo {author} {\bibfnamefont {P.}~\bibnamefont {He}}, \bibinfo
  {author} {\bibfnamefont {B.~A.}\ \bibnamefont {Olsen}}, \bibinfo {author}
  {\bibfnamefont {K.~G.}\ \bibnamefont {Jackson}}, \bibinfo {author}
  {\bibfnamefont {H.}~\bibnamefont {Sharum}}, \bibinfo {author} {\bibfnamefont
  {S.}~\bibnamefont {Trotzky}}, \bibinfo {author} {\bibfnamefont
  {J.}~\bibnamefont {Marino}}, \bibinfo {author} {\bibfnamefont {A.~M.}\
  \bibnamefont {Rey}},\ and\ \bibinfo {author} {\bibfnamefont {J.~H.}\
  \bibnamefont {Thywissen}},\ }\bibfield  {title} {\bibinfo {title}
  {Observation of a transition between dynamical phases in a quantum degenerate
  {Fermi} gas},\ }\href@noop {} {\bibfield  {journal} {\bibinfo  {journal}
  {Science Advances}\ }\textbf {\bibinfo {volume} {5}} (\bibinfo {year}
  {2019})},\ \bibinfo {note} {elocation-id: eaax1568}\BibitemShut {NoStop}%
\bibitem [{\citenamefont {Koller}\ \emph {et~al.}(2016)\citenamefont {Koller},
  \citenamefont {Wall}, \citenamefont {Mundinger},\ and\ \citenamefont
  {Rey}}]{KollerReySpinDep}%
  \BibitemOpen
  \bibfield  {author} {\bibinfo {author} {\bibfnamefont {A.~P.}\ \bibnamefont
  {Koller}}, \bibinfo {author} {\bibfnamefont {M.~L.}\ \bibnamefont {Wall}},
  \bibinfo {author} {\bibfnamefont {J.}~\bibnamefont {Mundinger}},\ and\
  \bibinfo {author} {\bibfnamefont {A.~M.}\ \bibnamefont {Rey}},\ }\bibfield
  {title} {\bibinfo {title} {Dynamics of interacting fermions in spin-dependent
  potentials},\ }\href@noop {} {\bibfield  {journal} {\bibinfo  {journal}
  {Phys. Rev. Lett.}\ }\textbf {\bibinfo {volume} {117}},\ \bibinfo {pages}
  {195302} (\bibinfo {year} {2016})}\BibitemShut {NoStop}%
\bibitem [{\citenamefont {Wall}(2020)}]{WallEnergySpinLattice}%
  \BibitemOpen
  \bibfield  {author} {\bibinfo {author} {\bibfnamefont {M.~L.}\ \bibnamefont
  {Wall}},\ }\bibfield  {title} {\bibinfo {title} {Simulating fermions in
  spin-dependent potentials with spin models on an energy lattice},\
  }\href@noop {} {\bibfield  {journal} {\bibinfo  {journal} {Phys. Rev. A}\
  }\textbf {\bibinfo {volume} {102}},\ \bibinfo {pages} {023329} (\bibinfo
  {year} {2020})}\BibitemShut {NoStop}%
\bibitem [{Sup()}]{SupportOnline}%
  \BibitemOpen
  \bibinfo {note} {See the supplementary material for a description of the
  experimental methods and a detailed discussion of the spin evolution model
  including light-induced loss.}\BibitemShut {Stop}%
\bibitem [{\citenamefont {Huang}\ and\ \citenamefont
  {Thomas}(2024)}]{JingjingCorrel}%
  \BibitemOpen
  \bibfield  {author} {\bibinfo {author} {\bibfnamefont {J.}~\bibnamefont
  {Huang}}\ and\ \bibinfo {author} {\bibfnamefont {J.~E.}\ \bibnamefont
  {Thomas}},\ }\bibfield  {title} {\bibinfo {title} {Energy-resolved spin
  correlation measurements: Decoding transverse spin dynamics in weakly
  interacting {Fermi} gases},\ }\href
  {https://doi.org/10.1103/PhysRevA.109.L041301} {\bibfield  {journal}
  {\bibinfo  {journal} {Phys. Rev. A}\ }\textbf {\bibinfo {volume} {109}},\
  \bibinfo {pages} {L041301} (\bibinfo {year} {2024})}\BibitemShut {NoStop}%
\bibitem [{\citenamefont {Margalit}\ \emph {et~al.}(2021)\citenamefont
  {Margalit}, \citenamefont {Lu}, \citenamefont {Çağrı Top},\ and\
  \citenamefont {Ketterle}}]{PauliKetterleHighDensity}%
  \BibitemOpen
  \bibfield  {author} {\bibinfo {author} {\bibfnamefont {Y.}~\bibnamefont
  {Margalit}}, \bibinfo {author} {\bibfnamefont {Y.-K.}\ \bibnamefont {Lu}},
  \bibinfo {author} {\bibfnamefont {F.}~\bibnamefont {Çağrı Top}},\ and\
  \bibinfo {author} {\bibfnamefont {W.}~\bibnamefont {Ketterle}},\ }\bibfield
  {title} {\bibinfo {title} {Pauli blocking of light scattering in degenerate
  fermions},\ }\href@noop {} {\bibfield  {journal} {\bibinfo  {journal}
  {Science}\ }\textbf {\bibinfo {volume} {374}},\ \bibinfo {pages} {976}
  (\bibinfo {year} {2021})}\BibitemShut {NoStop}%
\bibitem [{\citenamefont {Sanner}\ \emph {et~al.}(2021)\citenamefont {Sanner},
  \citenamefont {Sonderhouse}, \citenamefont {Hutson}, \citenamefont {Yan},
  \citenamefont {Milner},\ and\ \citenamefont {Ye}}]{PauliAtomLight}%
  \BibitemOpen
  \bibfield  {author} {\bibinfo {author} {\bibfnamefont {C.}~\bibnamefont
  {Sanner}}, \bibinfo {author} {\bibfnamefont {L.}~\bibnamefont {Sonderhouse}},
  \bibinfo {author} {\bibfnamefont {R.~B.}\ \bibnamefont {Hutson}}, \bibinfo
  {author} {\bibfnamefont {L.}~\bibnamefont {Yan}}, \bibinfo {author}
  {\bibfnamefont {W.~R.}\ \bibnamefont {Milner}},\ and\ \bibinfo {author}
  {\bibfnamefont {J.}~\bibnamefont {Ye}},\ }\bibfield  {title} {\bibinfo
  {title} {Pauli blocking of atom-light scattering},\ }\href@noop {} {\bibfield
   {journal} {\bibinfo  {journal} {Science}\ }\textbf {\bibinfo {volume}
  {374}},\ \bibinfo {pages} {979} (\bibinfo {year} {2021})}\BibitemShut
  {NoStop}%
\bibitem [{\citenamefont {Deb}\ and\ \citenamefont
  {{Kj{æ}rgaard}}(2021)}]{PauliLightScatt}%
  \BibitemOpen
  \bibfield  {author} {\bibinfo {author} {\bibfnamefont {A.~B.}\ \bibnamefont
  {Deb}}\ and\ \bibinfo {author} {\bibfnamefont {N.}~\bibnamefont
  {{Kj{æ}rgaard}}},\ }\bibfield  {title} {\bibinfo {title} {Observation of
  {Pauli} blocking in light scattering from quantum degenerate fermions},\
  }\href@noop {} {\bibfield  {journal} {\bibinfo  {journal} {Science}\ }\textbf
  {\bibinfo {volume} {374}},\ \bibinfo {pages} {972} (\bibinfo {year}
  {2021})}\BibitemShut {NoStop}%
\bibitem [{\citenamefont {Jannin}\ \emph {et~al.}(2022)\citenamefont {Jannin},
  \citenamefont {van~der Werf}, \citenamefont {Steinebach}, \citenamefont
  {Bethlem},\ and\ \citenamefont {Eikema}}]{PauliStimEm}%
  \BibitemOpen
  \bibfield  {author} {\bibinfo {author} {\bibfnamefont {R.}~\bibnamefont
  {Jannin}}, \bibinfo {author} {\bibfnamefont {Y.}~\bibnamefont {van~der
  Werf}}, \bibinfo {author} {\bibfnamefont {K.}~\bibnamefont {Steinebach}},
  \bibinfo {author} {\bibfnamefont {H.~L.}\ \bibnamefont {Bethlem}},\ and\
  \bibinfo {author} {\bibfnamefont {K.~S.~E.}\ \bibnamefont {Eikema}},\
  }\bibfield  {title} {\bibinfo {title} {Pauli blocking of stimulated emission
  in a degenerate {Fermi} gas},\ }\href@noop {} {\bibfield  {journal} {\bibinfo
   {journal} {Nature Communications}\ }\textbf {\bibinfo {volume} {13}},\
  \bibinfo {pages} {6479} (\bibinfo {year} {2022})}\BibitemShut {NoStop}%
\bibitem [{\citenamefont {Arunkumar}\ \emph {et~al.}(2019)\citenamefont
  {Arunkumar}, \citenamefont {Jagannathan},\ and\ \citenamefont
  {Thomas}}]{NithyaSpatial}%
  \BibitemOpen
  \bibfield  {author} {\bibinfo {author} {\bibfnamefont {N.}~\bibnamefont
  {Arunkumar}}, \bibinfo {author} {\bibfnamefont {A.}~\bibnamefont
  {Jagannathan}},\ and\ \bibinfo {author} {\bibfnamefont {J.~E.}\ \bibnamefont
  {Thomas}},\ }\bibfield  {title} {\bibinfo {title} {Designer spatial control
  of interactions in ultracold gases},\ }\href@noop {} {\bibfield  {journal}
  {\bibinfo  {journal} {Phys. Rev. Lett.}\ }\textbf {\bibinfo {volume} {122}},\
  \bibinfo {pages} {040405} (\bibinfo {year} {2019})}\BibitemShut {NoStop}%
\bibitem [{\citenamefont {Hazzard}\ and\ \citenamefont
  {Gadway}(2023)}]{synthLattice}%
  \BibitemOpen
  \bibfield  {author} {\bibinfo {author} {\bibfnamefont {K.}~\bibnamefont
  {Hazzard}}\ and\ \bibinfo {author} {\bibfnamefont {B.}~\bibnamefont
  {Gadway}},\ }\bibfield  {title} {\bibinfo {title} {Synthetic dimensions},\
  }\href@noop {} {\bibfield  {journal} {\bibinfo  {journal} {Physics Today}\
  }\textbf {\bibinfo {volume} {76(4)}},\ \bibinfo {pages} {62} (\bibinfo {year}
  {2023})}\BibitemShut {NoStop}%
\bibitem [{\citenamefont {Huang}\ \emph {et~al.}(2023)\citenamefont {Huang},
  \citenamefont {Royse}, \citenamefont {Arakelyan},\ and\ \citenamefont
  {Thomas}}]{JingjingRewind}%
  \BibitemOpen
  \bibfield  {author} {\bibinfo {author} {\bibfnamefont {J.}~\bibnamefont
  {Huang}}, \bibinfo {author} {\bibfnamefont {C.~A.}\ \bibnamefont {Royse}},
  \bibinfo {author} {\bibfnamefont {I.}~\bibnamefont {Arakelyan}},\ and\
  \bibinfo {author} {\bibfnamefont {J.~E.}\ \bibnamefont {Thomas}},\ }\bibfield
   {title} {\bibinfo {title} {Verifying a quasiclassical spin model of
  perturbed quantum rewinding in a {Fermi} gas},\ }\href@noop {} {\bibfield
  {journal} {\bibinfo  {journal} {Phys. Rev. A}\ }\textbf {\bibinfo {volume}
  {108}},\ \bibinfo {pages} {L041304} (\bibinfo {year} {2023})}\BibitemShut
  {NoStop}%
\bibitem [{\citenamefont {Wu}\ and\ \citenamefont
  {Thomas}(2012{\natexlab{b}})}]{WuThomasEffRange}%
  \BibitemOpen
  \bibfield  {author} {\bibinfo {author} {\bibfnamefont {H.}~\bibnamefont
  {Wu}}\ and\ \bibinfo {author} {\bibfnamefont {J.~E.}\ \bibnamefont
  {Thomas}},\ }\bibfield  {title} {\bibinfo {title} {Optical control of the
  scattering length and effective range for magnetically tunable {Feshbach}
  resonances in ultracold gases},\ }\href@noop {} {\bibfield  {journal}
  {\bibinfo  {journal} {Phys. Rev. A}\ }\textbf {\bibinfo {volume} {86}},\
  \bibinfo {pages} {063625} (\bibinfo {year} {2012}{\natexlab{b}})}\BibitemShut
  {NoStop}%
\bibitem [{\citenamefont {Z\"urn}\ \emph {et~al.}(2013)\citenamefont {Z\"urn},
  \citenamefont {Lompe}, \citenamefont {Wenz}, \citenamefont {Jochim},
  \citenamefont {Julienne},\ and\ \citenamefont
  {Hutson}}]{JochimPreciseFeshbach}%
  \BibitemOpen
  \bibfield  {author} {\bibinfo {author} {\bibfnamefont {G.}~\bibnamefont
  {Z\"urn}}, \bibinfo {author} {\bibfnamefont {T.}~\bibnamefont {Lompe}},
  \bibinfo {author} {\bibfnamefont {A.~N.}\ \bibnamefont {Wenz}}, \bibinfo
  {author} {\bibfnamefont {S.}~\bibnamefont {Jochim}}, \bibinfo {author}
  {\bibfnamefont {P.~S.}\ \bibnamefont {Julienne}},\ and\ \bibinfo {author}
  {\bibfnamefont {J.~M.}\ \bibnamefont {Hutson}},\ }\bibfield  {title}
  {\bibinfo {title} {Precise characterization of $^{6}\mathrm{Li}$ {Feshbach}
  resonances using trap-sideband-resolved rf spectroscopy of weakly bound
  molecules},\ }\href@noop {} {\bibfield  {journal} {\bibinfo  {journal} {Phys.
  Rev. Lett.}\ }\textbf {\bibinfo {volume} {110}},\ \bibinfo {pages} {135301}
  (\bibinfo {year} {2013})}\BibitemShut {NoStop}%
\bibitem [{\citenamefont {Bratten}\ and\ \citenamefont
  {Hammer}(2006)}]{BraatenReview}%
  \BibitemOpen
  \bibfield  {author} {\bibinfo {author} {\bibfnamefont {E.}~\bibnamefont
  {Bratten}}\ and\ \bibinfo {author} {\bibfnamefont {H.-W.}\ \bibnamefont
  {Hammer}},\ }\bibfield  {title} {\bibinfo {title} {Universality in few-body
  systems with large scattering length}} (\bibinfo {year} {2006}),\ \bibinfo
  {note} {arxiv.org/abs/cond-mat/0410417v3}\BibitemShut {NoStop}%
\bibitem [{\citenamefont {Bartenstein}\ \emph {et~al.}(2005)\citenamefont
  {Bartenstein}, \citenamefont {Altmeyer}, \citenamefont {Riedl}, \citenamefont
  {Geursen}, \citenamefont {Jochim}, \citenamefont {Chin}, \citenamefont
  {Denschlag}, \citenamefont {Grimm}, \citenamefont {Simoni}, \citenamefont
  {Tiesinga}, \citenamefont {Williams},\ and\ \citenamefont
  {Julienne}}]{BartensteinFeshbach}%
  \BibitemOpen
  \bibfield  {author} {\bibinfo {author} {\bibfnamefont {M.}~\bibnamefont
  {Bartenstein}}, \bibinfo {author} {\bibfnamefont {A.}~\bibnamefont
  {Altmeyer}}, \bibinfo {author} {\bibfnamefont {S.}~\bibnamefont {Riedl}},
  \bibinfo {author} {\bibfnamefont {R.}~\bibnamefont {Geursen}}, \bibinfo
  {author} {\bibfnamefont {S.}~\bibnamefont {Jochim}}, \bibinfo {author}
  {\bibfnamefont {C.}~\bibnamefont {Chin}}, \bibinfo {author} {\bibfnamefont
  {J.~H.}\ \bibnamefont {Denschlag}}, \bibinfo {author} {\bibfnamefont
  {R.}~\bibnamefont {Grimm}}, \bibinfo {author} {\bibfnamefont
  {A.}~\bibnamefont {Simoni}}, \bibinfo {author} {\bibfnamefont
  {E.}~\bibnamefont {Tiesinga}}, \bibinfo {author} {\bibfnamefont {C.~J.}\
  \bibnamefont {Williams}},\ and\ \bibinfo {author} {\bibfnamefont {P.~S.}\
  \bibnamefont {Julienne}},\ }\bibfield  {title} {\bibinfo {title} {Precise
  determination of $^6$\mbox{Li} cold collision parameters by radio-frequency
  spectroscopy on weakly bound molecules},\ }\href@noop {} {\bibfield
  {journal} {\bibinfo  {journal} {Phys. Rev. Lett.}\ }\textbf {\bibinfo
  {volume} {94}},\ \bibinfo {pages} {103201} (\bibinfo {year}
  {2005})}\BibitemShut {NoStop}%
\end{thebibliography}
\end{document}